\documentclass[a4paper,11pt]{article}
\pdfoutput=1 

\usepackage{jheppub} 
\usepackage[usenames]{color}
\usepackage[T1]{fontenc} 
\usepackage{slashed}
\usepackage{amsmath}
\usepackage{amssymb}

\title{\boldmath Lepton Flavour Violation in Hadron Decays of the Tau Lepton within the Littlest Higgs Model with T-parity}

\author{Iván Pacheco}
\author{and Pablo Roig}
\affiliation{Departamento de Física, Centro de Investigación y de Estudios Avanzados del Instituto Politécnico Nacional \\
Apartado Postal 14-740, 07000 Ciudad de México, México}

\emailAdd{ipacheco@fis.cinvestav.mx}
\emailAdd{proig@fis.cinvestav.mx}

\abstract{We first study the hadronic lepton flavor violating tau decays within the littlest Higgs model with T-parity (including one or two pseudoscalars, or a vector resonance). We consider the case where only T-odd particles and partner fermions contribute, and also its extension including Majorana neutrinos coming from an inverse seesaw. In both cases our mean values lie mostly only one order of magnitude below current upper limits, strengthening the case of searching for these decays in the quest for new physics.}


\arxivnumber{2207.04085}

\begin{document} 
\maketitle

\section{Introduction}
\label{sec:Intro}
A scalar boson mass is not protected by any symmetry, unlike the case of spin $1/2$ fermions and spin $1$ bosons. For them, chiral and gauge symmetries, respectively, ensure that the quantum corrections to their masses vanish in the limit of small bare masses. The presence of a Higgs boson \cite{Higgs:1964pj, Higgs:1964ia,Englert:1964et, Guralnik:1964eu} with mass on the electroweak scale ($\sim v$) \cite{ATLAS:2012yve, CMS:2012qbp} is unnatural upon the existence of heavy new particles coupling to it proportionally to their masses. This was christened as the hierarchy problem \cite{Georgi:1974yf}, whose solution is still unknown. The fact that the Standard Model (SM) \cite{Glashow:1961tr, Weinberg:1967tq, Salam:1968rm} is incomplete~\footnote{At least it lacks a mechanism to generate the observed baryon asymmetry of the universe.}, together with the strong restrictions on new physics close to $v$ coming from very accurate low-energy measurements, electroweak precision data and LHC constraints \cite{PDG}, aggravates the hierarchy issue.  Famous solutions to it include supersymmetry or extra dimensions, and little Higgs models, to which our work belongs.

Composite Higgs models \cite{Arkani-Hamed:2002ikv, Schmaltz:2005ky, Perelstein:2005ka, Panico:2015jxa} update the ideas of technicolor \cite{Weinberg:1975gm}. Thus, the Higgs boson is a pseudo-Nambu-Goldstone boson  of a 
spontaneously broken global symmetry at a scale $f>v$. As a result, the Higgs mass is of $\mathcal{O}(v)$ despite the existence of new physics at the scale $f$. In this way the hierarchy problem is not solved but delayed up to the energy where the model becomes strongly coupled. Specifically, we will work within the Littlest Higgs model with T-parity (LHT) \cite{Arkani-Hamed:2002ikv, Arkani-Hamed:2001kyx, Arkani-Hamed:2001nha, Cheng:2003ju, Cheng:2004yc, Low:2004xc, Cheng:2005as}, whose fermion sector is reviewed in section \ref{sec:LHT}. The spontaneous collective breaking of the LHT corresponds to a 
global symmetry group $SU(5)$ broken down to $SO(5)$, by a vacuum expectation value at the TeV scale. Discrete T parity symmetry is imposed to avoid singly-produced new heavy particles and tree level corrections to SM observables. Thus, LHT is not stringently constrained by data \cite{Hubisz:2005tx, Hubisz:2005bd}. Varied aspects of its phenomenology have been studied in refs.~\cite{Hubisz:2005tx, Hubisz:2005bd, Chen:2006cs, Blanke:2006sb, Buras:2006wk, Belyaev:2006jh, Blanke:2006eb, Hill:2007zv, Goto:2008fj, Blanke:2009am, Han:2013ic, Yang:2013lpa, Reuter:2013iya, Yang:2014mba, Blanke:2015wba, Dercks:2018hgz, Illana:2021uwu}, in particular lepton flavor violating processes have been analyzed in refs.~\cite{Blanke:2007db,delAguila:2008zu, Blanke:2009am, delAguila:2010nv, Goto:2010sn,Liu:2010wav,Ma:2010gv, Han:2011zza,Goto:2015iha, Yang:2016hrh, delAguila:2017ugt, delAguila:2019htj, DelAguila:2019xec}.

LHT possibly realizing a low-scale seesaw and in particular one of inverse type (ISS) was put forward recently in ref.~\cite{DelAguila:2019xec}. In our previous article \cite{Pacheco}, we showed that purely leptonic LFV processes and $\mu\to e$ conversion in nuclei were closer to current bounds within this setting than in the standard LHT (see e.g. ref.~\cite{delAguila:2019htj}). These results motivated us to analyze semileptonic LFV tau decays in this work. Although they are quite suppressed in the Simplest Little Higgs model (SLH) \cite{Lami}, the richer composition of the LHT (even in absence of additional Majorana neutrinos coming from the ISS) should enable larger rates than in the SLH. We will show that branching ratios within one order of magnitude of current upper limits can be expected in the processes studied in this work.

This article is structured as follows: after a brief review of the fermion sector of the LHT in section \ref{sec:LHT}, we summarize the realization of the ISS within the LHT in section \ref{sec:inverseseesaw}. Then, in section \ref{sec:LFV} we present the different contributions to the considered semileptonic LFV tau decays. We start with the parton process in section \ref{sec:lqq} and continue -accounting for the corresponding hadronization. Associated phenomenology is discussed in section \ref{sec:Pheno} (first in the standard LHT and then adding Majorana neutrino contributions, arising from the ISS) and our conclusions are given in section \ref{sec:Concl}. Appendices include form factors for $T$-odd (appendix \ref{App: A}) and Majorana (appendix \ref{App: B}) contributions as well as a survey on hadronization techniques (appendix \ref{app:C}) and corresponding form factors (appendix \ref{app:D}).

\section{Littlest Higgs model with T parity (LHT)} 
\label{sec:LHT}
We develop here just the fermion sector to illustrate what are the new particles in the LHT and how they acquire mass. 
We embed two $SU(5)$ incomplete quintuplets and introduce a right-handed $SO(5)$ multiplet $\Psi_{R}$ \cite{Monika Blanke-066}
\begin{equation}
    \Psi_{1} = \left( \begin{array}{c}
         i\psi_{1}\\ 0 \\ 0
    \end{array} \right), \quad \Psi_{2} = \left( \begin{array}{c}
         0\\ 0 \\i\psi_{2}
    \end{array} \right), \quad \Psi_{R} = \left( \begin{array}{c}
        \tilde{\psi}^{c}_{R} \\ \chi_{R}\\ \psi_{R}
    \end{array} \right),
    \label{eq7.1.19}
\end{equation}
where the tilde denotes a partner lepton field, not to be confused with charge conjugation, and $\chi_{R}$ is a lepton singlet (see section \ref{sec:inverseseesaw}). Above ($\sigma^{2}$ is the second Pauli matrix)
\begin{equation}
    \psi_{i} = 
    - \sigma^{2}\left( \begin{array}{c}
          \nu_{L_{i}} \\ \ell_{L_{i}}
    \end{array} \right) \qquad (i = 1,2), \qquad \psi_{R} = -i \sigma^{2}\left( \begin{array}{c}
          \nu_{HR} \\ \ell_{HR}
    \end{array} \right).  
    \label{eq7.1.20}
\end{equation}
The action of T-parity on the LH(RH) leptons is then defined to be
\begin{equation}
    \Psi_{1} \longleftrightarrow \Omega \Sigma_{0}\Psi_{2}, \qquad \Psi_{R} \to \Omega \Psi_{R},
    \label{eq179}
\end{equation}
with
\begin{equation}
    \Omega = \mathrm{diag}(-1,-1,1,-1,-1), \quad \Sigma_{0} = \left(\begin{array}{ccc}
        0 & 0 & \textbf{1}_{2 \times 2} \\0&1&0\\
        \textbf{1}_{2 \times 2}&0&0
    \end{array}\right).
    \label{eq180}
\end{equation}
This discrete symmetry is implemented in the fermion sector duplicating the SM doublet $l_{L} = (l_{1L}-l_{2L})/\sqrt{2}$, corresponding to the T-even combination $(\Psi_{1}+\Omega \Sigma_{0} \Psi_{2})/\sqrt{2}$ , that remains light, with an extra heavy mirror doublet $l_{HL} = \left( \nu_{HL} ,\, \ell_{HL} \right)^{T} = \left( l_{1L}+l_{2L} \right)/\sqrt{2}$ obtained from the T-odd orthogonal combination $(\Psi_{1}-\Omega \Sigma_{0} \Psi_{2})/\sqrt{2}$. The mirror leptons can be given $\mathcal{O}(f)$ masses via \cite{DelAguila:2019xec}
\begin{equation}
    \mathcal{L}_{Y_{H}} = - \kappa f \left( \overline{\Psi}_{2} \xi + \overline{\Psi}_{1} \Sigma_{0} \xi^{\dagger} \right) \Psi_{R} + h.c.,
    \label{eq7.1.22}
\end{equation}
where $\xi = e^{i\Pi/f}$ and $f$ is the new physics (NP) energy scale of $\mathcal{O}$(TeV). For $\xi = \exp{\left(i\Pi/f\right)} \approx \mathbb{I}$, this Lagrangian gives a vector-like mass $\sqrt{2}\kappa f$ to $\nu_{H}$. The mirror leptons thus acquire masses after EWSB, given by \cite{Monika Blanke-066, Jay Hubisz-041}
\begin{equation}
     m_{\ell_{H}^{i}} = \sqrt{2}\kappa_{ii}f \equiv m_{Hi}, \quad m_{\nu^{i}_{H}} = m_{Hi} \left( 1 - \frac{\upsilon^{2}}{8f^{2}} \right),
     \label{eq7.1.23}
\end{equation}
where $\kappa_{ii}$ are the eigenvalues of the mass matrix $\kappa$. Partner leptons $\tilde{\ell}^{c} = (\tilde{\nu}^{c}, \tilde{\ell}^{c})$ also appear. To give them masses we need to introduce two incomplete $SO(5)$ multiplets \cite{delAguila:2017ugt, Jay Hubisz and Patrick Meade, Ian Low-067}
\begin{equation}
     \Psi_{L} = \left( \begin{array}{c}
        \tilde{\psi}^{c}_{L} \\ 0 \\ 0
    \end{array} \right), \quad \Psi_{L}^{\chi} = \left( \begin{array}{c}
        0 \\ \chi_{L} \\ 0
    \end{array} \right), \quad \mathrm{with} \ \Psi_{L} \xrightarrow{T} \Omega \Psi_{L}, \ \mathrm{and} \ \Psi_{L}^{\chi} \xrightarrow{T} \Omega \Psi_{L}^{\chi}.
    \label{eq7.1.0}
\end{equation}
These multiplets induct Dirac mass terms for the $\tilde{\psi}^{c}_{R}$ and $\chi_{R}$ fields as follows
\begin{align}
     \mathcal{L}_{Y_{H}} & = - \kappa f \left( \overline{\Psi}_{2} \xi + \overline{\Psi}_{1} \Sigma_{0} \xi^{\dagger} \right) \Psi_{R} - \kappa_{2}\overline{\Psi_{L}} \Psi_{R} - M \overline{\Psi_{L}^{\chi}}\Psi_{R} + \mathrm{h.c.} \nonumber \\
     & = - \kappa f \left( \overline{\Psi}_{2} \xi + \overline{\Psi}_{1} \Sigma_{0} \xi^{\dagger} \right) \Psi_{R} - \kappa_{2}\overline{\tilde{\psi}^{c}_{L}} \tilde{\psi}^{c}_{R} - M \overline{\chi_{L}}\chi_{R} + \mathrm{h.c.}
    \label{eq7.1.01}
\end{align}
Thus, partner leptons and singlets receive  $\kappa_{2}$ and $M$ masses, respectively. 

It happens similarly for the masses of T-odd quarks, with $\kappa_{ii}^{q}$ instead of $\kappa_{ii}$, and replacing $d_{H}$-quark by $\ell_{H}$, $u_{H}$-quark by $\nu_{H}$, and $\kappa_{2}^{q}$ by $\kappa_{2}$ for the mass matrix of partner quarks of $u$ and $d$ types.

\section{Inverse seesaw neutrino masses in the LHT model}\label{sec:inverseseesaw}

The lepton singlets $\chi_{R}$  also get a large (vector–like) mass by combining with a LH singlet $\chi_{L}$ through a direct mass term without further couplings to the Higgs. Therefore, its mass term is written (as the eq. (\ref{eq7.1.01}))
\begin{equation}
    \mathcal{L}_{M} = - M \overline{\chi}_{L}\chi_{R} + h.c.
    \label{eq184}
\end{equation}
We stress that $\chi_{L}$ is an $SU(5)$ singlet. Consequently, it is natural to include a small Majorana mass for it. Once lepton number is assumed to be only broken by small Majorana masses $\mu$ in the heavy LH neutral sector,
\begin{equation}
    \mathcal{L}_{\mu} = - \frac{\mu}{2} \overline{\chi^{c}_{L}}\chi_{L} + h.c.,
    \label{eq185}
\end{equation}
so the resulting (T-even) neutrino mass matrix reduces to the inverse see-saw one \cite{DelAguila:2019xec}
\begin{equation}
    \mathcal{L}_{M}^{\nu} = - \frac{1}{2} \left( \overline{\nu^{c}_{l}} \ \overline{\chi_{R}} \ \overline{\chi^{c}_{L}} \right) \mathcal{M}_{\nu}^{T-even} \left( \begin{array}{c}
         \nu_{L} \\ \chi_{R}^{c} \\ \chi_{L}
    \end{array} \right) + h.c.,
    \label{eq186}
\end{equation}
where 
\begin{equation}
    \mathcal{M}_{\nu}^{T-even} = \left( \begin{array}{ccc}
         0 & i \kappa^{*}f sin \left( \frac{\upsilon}{\sqrt{2}f} \right) & 0\\
        i \kappa^{\dagger}f sin \left( \frac{\upsilon}{\sqrt{2}f} \right) & 0 & M^{\dagger} \\
        0 & M^{*} & \mu
    \end{array} \right),
    \label{eq187}
\end{equation}
with each entry standing for a $3 \times 3$ matrix to account for the 3 lepton families. The $\kappa$ entries are given by the 
Lagrangian in eq. (\ref{eq7.1.01}) and $M$ stands for the direct heavy Dirac mass matrix from eq. (\ref{eq184}). Finally, $\mu$ is the mass matrix of small Majorana masses in eq. (\ref{eq185}).

Upon considering the hierarchy $\mu \ll \kappa f \ll M$ (inverse see-saw), the mass eigenvalues for $M$ are $\sim 10$ TeV, of the order of $4\pi f$ with $f \sim$ TeV, as required by current EWPD if we assume the $\kappa$ eigenvalues to be order $1$ \cite{Pacheco}. Conversely,  the $\mu$ eigenvalues shall be much smaller than the GeV.

After diagonalizing the mass matrix from eq.(\ref{eq187}), in the mass eigenstates basis, the SM charged and neutral currents become \cite{Pacheco}
\begin{equation}
\begin{split}
   \mathcal{L}_{W}^{l} & = \frac{g}{\sqrt{2}} W_{\mu}^{+} \sum_{i,j=1}^{3} \overline{\nu_{i}^{l}} W_{ij}\gamma^{\mu}P_{L} \ell_{j} + h.c., \quad \mathrm{with} \quad W_{ij} = \sum_{k=1}^{3}  U^{\dagger}_{ik} [\mathbf{1}_{3 \times 3} - \frac{1}{2}(\theta \theta^{\dagger})]_{kj},\\
   \mathcal{L}_{W}^{lh} & = \frac{g}{\sqrt{2}} W_{\mu}^{+} \sum_{i,j=1}^{3}  \overline{\chi_{i}^{h}} \theta^{\dagger}_{ij} \gamma^{\mu}P_{L} \ell_{j} + h.c.
\end{split}
\label{eq211}
\end{equation}
and
\begin{equation}
    \begin{split}
        \mathcal{L}_{Z}^{l} & = \frac{g}{2 \cos\theta_{W}} Z_{\mu} \sum_{i,j=1}^{3} \overline{\nu_{i}^{l}} \gamma^{\mu}(X_{ij}P_{L}-X_{ij}^{\dagger}P_{R}) \nu_{j}^{l}, \quad \mathrm{with} \quad X_{ij} = \sum_{k=1}^{3} \left(U^{\dagger} [\mathbf{1}_{3 \times 3} - (\theta \theta^{\dagger})]\right)_{ik} U_{kj}, \\
        \mathcal{L}_{Z}^{lh} & = \frac{g}{2cos\theta_{W}} Z_{\mu} \sum_{i,j=1}^{3} \overline{\chi_{i}^{h}} \gamma^{\mu} (Y_{ij}P_{L}-Y_{ij}^{\dagger}P_{R})\nu_{j}^{l} + h.c.,\quad \mathrm{with} \quad Y_{ij} = \sum_{k=1}^{3} \theta^{\dagger}_{ik} U_{kj}, \\
        \mathcal{L}_{Z}^{h} & = \frac{g}{2\cos\theta_{W}} Z_{\mu} \sum_{i,j=1}^{3} \overline{\chi_{i}^{h}}\gamma^{\mu}(S_{ij}P_{L}-S_{ij}^{\dagger}P_{R})\chi_{j}^{h},\quad \mathrm{with} \quad S_{ij} = \sum_{k=1}^{3} \theta^{\dagger}_{ik} \theta_{kj}\,,
    \end{split}
    \label{eq212}
\end{equation}
where $\theta$ matrix elements give the mixing between light and heavy (quasi-Dirac) neutrinos to leading order, and $U = U_{PMNS}$.

Assuming universality and, in particular, that the three
mixing angles $\theta_{ii}$ are equal, their absolute value is found to be $< 0.03 \  \mathrm{at} \ 95\%$ C.L. \cite{JdeBlas}. Hence,
\begin{equation}
    |\kappa_{ii}| < 0.17 \left( \frac{\mathrm{M}_{i}}{\mathrm{TeV}} \right) \leq 2.136 \left( \frac{f}{\mathrm{TeV}} \right),
    \label{eq7.6.3}
\end{equation}
for $f>1$ TeV. This effective description requires $\mathrm{M}_{i} \lesssim 4\pi f$, with $\mathrm{M}_{i}$ the heavy Majorana neutrino masses.

\section{Lepton Flavour Violating Hadron Decays of the Tau Lepton}
\label{sec:LFV}

In this section we apply our model for the study of LFV tau decays into hadrons: $\tau \to \mu P$, $\tau \to \mu V$ and $\tau \to  \mu P\bar{P}$ where $P \ (V)$ is short for a pseudoscalar (vector) meson~\footnote{Effective field theory analyses of these processes can be found in refs.~\cite{Celis:2013xja, Celis:2014asa, Husek:2020fru,Cirigliano:2021img}.}. For the single meson case, we will consider $P=\pi^0,\,\eta,\,\eta'$. For the two-meson channels, we will restrict ourselves to the numerically leading $P=\pi,\,K$ cases. Correspondingly, $V=\rho,\phi$ ($\omega$ being suppressed).\\
We are going to consider the effects of T-odd and partner particles, as well as heavy Majorana neutrinos, since terms of order  $\mathcal{O}(\upsilon^{2}/f^{2})$ are taken into account ($\upsilon$ is the vev of the SM Higgs and $f$ is the NP energy scale $\sim$ TeV, such that $\upsilon^{2}/f^{2} \ll 1$). We begin our analysis determining the amplitudes of the $\tau \to \mu q \bar{q}$ process, with $q = u,\, d,\, s$ quarks and, afterwards, proceed to hadronize the corresponding quarks bilinears. For this latter step we
will employ the tools given by chiral symmetry and dispersion relations, enforcing the right short-distance behaviour to the form factors.

In the following calculations  several mass ratios  will appear,  corresponding to the particles running in the loops. Thus, it will be useful to define the terminology from here on, as displayed in Table \ref{tab:terminology}.
\begin{table}[!ht]
    \centering
    \begin{tabular}{|c|c|}
    \hline
        Particles & Symbol \\
        \hline
        Mirror leptons & $x^{(\nu, \ell)_{H}} = m_{(\nu,\ell)_{H}}^{2}/M_{W_{H}}^{2}$ \\
        Partner leptons & $\tilde{x} = m_{\tilde{\nu}^{c}}^{2}/M_{\Phi}^{2}$ \\
        Mirror quarks & $y^{(u, d)_{H}} = m_{(u,d)_{H}}^{2}/M_{W_{H}}^{2}$ \\
        Partner quarks & $\tilde{y}^{\tilde{q}^{c}} = m_{(\tilde{u},\tilde{d} )^{c}}^{2}/M_{\Phi}^{2}$ \\
        SM quarks & $y_{i}^{q} = m_{q_{i}}^{2}/M_{W}^{2}$ \\
        Heavy Majorana neutrinos & $z_i = M_{W}^{2}/M_i^{2}$\\
        \hline
    \end{tabular}
    \caption{Terminology used in the following calculations, where we have assumed that $m_{\tilde{\nu}^{c}} = m_{\tilde{\ell}^{c}}$ and $M_i$ are the masses of the heavy Majorana neutrinos, that is the largest scale (which motivated our definition of the $z_i$ variable as $M_{W}^{2}/M_i^{2}$).}
    \label{tab:terminology}
\end{table}

\subsection{\texorpdfstring{$\tau \to \ell q \Bar{q} \ (\ell = e,\mu)$}{}}
\label{sec:lqq}

Two generic topologies are involved in this amplitude: i) penguin-like diagrams, namely $\tau \to \ell \{\gamma, Z\}$, followed by $\{\gamma, Z\} \to q\bar{q}$ and ii) box diagrams. We will assume, for simplicity, that light quarks and leptons ($\ell$) are massless in our calculation. Corrections induced by their finite masses can be safely neglected.\\
The full amplitude is given by two contributions: one of them comes from T-odd particles and the other from heavy Majorana neutrinos, then 
\begin{equation}
    \mathcal{M} = \mathcal{M}^{\mathrm{T-odd}} + \mathcal{M}^{\mathrm{Maj}},
    \label{eq7.1.1}
\end{equation}
where each one is written 
\begin{equation}
    \begin{split}
        \mathcal{M}^{\mathrm{T-odd}} & = \mathcal{M}^{\mathrm{T-odd}}_{\gamma} + \mathcal{M}^{\mathrm{T-odd}}_{Z} + \mathcal{M}^{\mathrm{T-odd}}_{\mathrm{box}}, \\
        \mathcal{M}^{\mathrm{Maj}} & = \mathcal{M}^{\mathrm{Maj}}_{\gamma} + \mathcal{M}^{\mathrm{Maj}}_{Z} + \mathcal{M}^{\mathrm{Maj}}_{\mathrm{box}},
    \end{split}
    \label{eq7.1.2}
\end{equation}
we will use the 't Hooft-Feynman gauge along the calculation and will write $\ell=\mu$ for definiteness.

\subsubsection{T-odd contribution}
\begin{figure}[!ht]
     \centering
     \includegraphics[scale = 0.65]{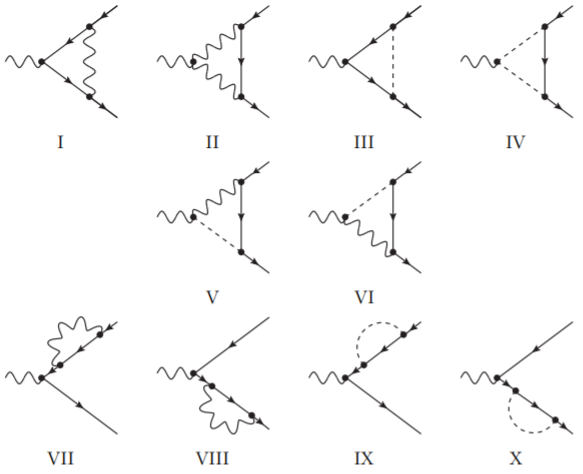}
     \caption{Topologies of the diagrams that contribute to the processes $\gamma,Z \rightarrow \Bar{\ell}\ell^{\prime}$.}
     \label{fig:topologies}
\end{figure}

The Feynman diagrams that contribute to the $\mathcal{M}^{\mathrm{T-odd}}_{\gamma}$ amplitude are shown in Figure \ref{fig:topologies}, whose structure  follows
\begin{equation}
    \begin{split}
         \mathcal{M}^{\mathrm{T-odd}}_{\gamma} & = \frac{e^{2}}{Q^{2}} \overline{\mu}(p') [2iF_{M}^{\gamma}(Q^{2})P_{R}\sigma^{\mu\nu}Q_{\nu} + F_{L}^{\gamma}(Q^{2})\gamma^{\mu}P_{L}]\tau(p) \overline{q}(p_{q}) Q_{q}\gamma_{\mu} q(p_{\bar{q}}),
     \end{split}
     \label{eq7.1.6}
\end{equation}
where $Q^{2} = (p_{q} + p_{\bar{q}})^{2}$ is the squared  momentum  transfer and $F_{M}^{\gamma}$ is given by~\footnote{Within the LHT, there appear new heavy gauge bosons and a scalar electroweak triplet with $\mathcal{O}(f)$ masses: $A_H,Z_H,W_H,\Phi$, see eq.~(\ref{eq7.6.1}). The corresponding Goldstone bosons of the former come from the $\Pi$ matrix \cite{delAguila:2019htj,delAguila:2008zu}, see eq.~(\ref{eq7.1.22}) and text below.}
\begin{equation}
    F_{M}^{\gamma} = F_{M}^{\gamma}|_{W_{H}} + F_{M}^{\gamma}|_{Z_{H}} + F_{M}^{\gamma}|_{A_{H}} + F_{M}^{\gamma}|_{\tilde{\nu}^{c}} + F_{M}^{\gamma}|_{\tilde{\ell}^{c}}. 
    \label{eq7.1.7}
\end{equation}

In this case we consider terms of order $\mathcal{O}(Q^{2})$. All functions below agree with ref.~\cite{delAguila:2019htj}. We write explicitly the above form factor
\begin{equation}
    \begin{split}
        F_{M}^{\gamma} = & \frac{\alpha_{W}}{16\pi}\frac{m_{\tau}}{M_{W}^{2}}\frac{\upsilon^{2}}{4f^{2}} \sum_{i} V_{H\ell}^{i\mu\dagger}V_{H\ell}^{i\tau} \left[ F_{W_{H}}(x_{i}^{\nu_{H}},Q^{2}) + F_{Z_{H}}(x_{i}^{\ell_{H}},Q^{2}) + \frac{1}{5}F_{Z_{H}}(ax_{i}^{\ell_{H}},Q^{2}) \right] \\
        & + \frac{\alpha_{W}}{16\pi} \frac{m_{\tau}}{M_{\Phi}^{2}} \frac{\upsilon^{2}}{4f^{2}} \sum_{i,j,k} V_{H \ell}^{i\mu \dagger}\frac{m_{\ell_{Hi}}}{M_{W}}W_{ij}^{\dagger}W_{jk}\frac{m_{\ell_{Hk}}}{M_{W}}V_{H\ell}^{k\tau} \left[F_{\tilde{\nu}^{c}}\left( \tilde{x}_{j},Q^{2} \right)+F_{\tilde{\ell}^{c}}\left( \tilde{x}_{j},Q^{2} \right)\right], 
    \end{split}
    \label{eq7.1.8}
\end{equation}
with $a = \frac{M_{W_{H}}^{2}}{M_{A_{H}}^{2}} = \frac{5c_{W}^{2}}{s^{2}_{W}}\sim15$ and
we have used $M_{W}^{2}/M_{W_{H}}^{2} = \upsilon^{2}/(4f^{2})$.
Each function above can be found in Appendix \ref{App: A}. Here $V_{H\ell}^{i}$ are the matrix elements of the $3 \times 3$ unitary mixing matrix parametrizing the misalignment between the SM left-handed charged leptons $\ell$ with the heavy mirror ones $\ell_{H}$. The $W_{jk}$ are analogous,  parametrizing the relative orientation of the mirror leptons and their partners $\tilde{\ell}^{c}$ in the $SO(5)$ (right-handed) multiplets \cite{delAguila:2019htj}.

$F_{L}^{\gamma}$ can be expressed as follows \cite{delAguila:2019htj, delAguila:2008zu}
\begin{equation}
    \begin{split}
        F_{L}^{\gamma} = & \frac{\alpha_{W}}{4\pi} \frac{Q^{2}}{M_{W_{H}}^{2}} \sum_{i}V_{H\ell}^{i\mu \dagger}V_{H\ell}^{i\tau} \left(G^{(1)}_{Z}(x_{i}^{\ell_{H}}) + \frac{1}{5}G_{Z}^{(1)}(ax_{i}^{\ell_{H}})+ G^{(1)}_{W}(x_{i}^{\nu_{H}}) \right) \\
        & + \frac{\alpha_{W}}{4\pi}\frac{Q^{2}}{2M_{h}^{2}} \frac{\upsilon^{4}}{4f^{4}}\sum_{ijk} V_{H \ell}^{i\mu \dagger}\frac{m_{\ell_{Hi}}}{M_{W}}W_{ij}^{\dagger}W_{jk}\frac{m_{\ell_{Hk}}}{M_{W}}V_{H\ell}^{k\tau}\left( G_{\tilde{\nu}^{c}}^{(1)}(\tilde{x}_{j}) + G_{\tilde{\ell}^{c}}^{(1)}(\tilde{x}_{j}) \right)\,,
    \end{split}
    \label{eq7.1.11}
\end{equation}
where $G_{Z}^{(1)}(x)$, $G_{W}^{(1)}(x)$, $G_{\tilde{\nu}^{c}}^{(1)}(x)$ and $G_{\tilde{\ell}^{c}}^{(1)}(x)$ are defined in Appendix \ref{App: A}.

The penguin-like diagrams with Z are given in Figure \ref{fig:topologies}. Their amplitude can be expressed
\begin{equation}
    \begin{split}
        \mathcal{M}_{Z}^{\mathrm{T-odd}} & = \frac{e^{2}}{M_{Z}^{2}} \overline{\mu}(p')[\gamma^{\mu} (F_{L}^{Z}P_{L} + F_{R}^{Z}P_{R})]\tau(p)\overline{q}(p_{q})[\gamma_{\mu} (Z_{L}P_{L} + Z_{R}P_{R})]q(p_{\bar{q}}),
    \end{split}
    \label{eq7.1.13}
\end{equation}
where
\begin{equation}
    \begin{split}
        Z_{L} & = \frac{g}{c_{W}} (T_{3}^{q} - s_{W}^{2}Q_{q}), \\
        Z_{R} & = -\frac{g}{c_{W}}s_{W}^{2}Q_{q},
    \end{split}
    \label{eq7.1.14}
\end{equation}
being
\begin{equation}
    T_{3}^{q} = \frac{1}{2} \left( \begin{array}{ccc}
        1 & & \\
        & -1 & \\
        & & -1
    \end{array} \right).
    \label{eq7.1.15}
\end{equation}
The corresponding right-handed vector form factor $F_{R}^{Z}$ is $\mathcal{O}\left( m_{\tau}^{2}/f^{2} \right)$ in the LHT and thus negligible as $f \sim \mathcal{O}$(TeV). The left-handed form factor $F_{L}^{Z}$, at order $\mathcal{O}(\mathrm{Q}^{2})$ according to \cite{delAguila:2019htj}, is given by
\begin{align}
    F_{L}^{Z} = & F_{L}^{Z}|_{W_{H}} + F_{L}^{Z}|_{A_{H}} + F_{L}^{Z}|_{Z_{H}}+F_{L}^{Z}|_{\tilde{\nu}^{c}}+F_{L}^{Z}|_{\tilde{\ell}^{c}} \nonumber \\
    = & \frac{\alpha_{W}}{8\pi c_{W}s_{W}} \sum_{i}V_{H\ell}^{i\mu\dagger}V_{H\ell}^{i\tau} \left\lbrace \frac{\upsilon^{2}}{8f^{2}} H_{L}^{W(0)}(x_{i}^{\nu_{H}}) \right. \nonumber \\
    & \left. + \frac{Q^{2}}{M_{W_{H}}^{2}} \left[ H_{L}^{W}(y_{i}) + (1-2c_{W}^{2}) \left( \frac{1}{5} H_{L}^{A/Z}(ax_{i}^{\ell_{H}}) + H_{L}^{A/Z}(x_{i}^{\ell_{H}}) \right) \right] \right\rbrace \nonumber \\
    & + \frac{\alpha_{W}}{8\pi c_{W}s_{W}}\frac{Q^{2}}{M_{h}^{2}} \frac{\upsilon^{2}}{2f^{2}}\sum_{ijk} V_{H \ell}^{i\mu \dagger}\frac{m_{\ell_{Hi}}}{M_{W_{H}}}W_{ij}^{\dagger}W_{jk}\frac{m_{\ell_{Hk}}}{M_{W_{H}}}V_{H\ell}^{k\tau} \left[ H_{L}^{\bar{\nu}}(\tilde{x}_{j}) + (1-2c_{W}^{2})H_{L}^{\bar{\ell}}(\tilde{x}_{j}) \right],
    \label{eq7.1.16}
\end{align}
where all  function  definitions can be found in Appendix \ref{App: A}.

\begin{figure}[!ht]
    \centering
    \includegraphics[scale = 0.4351]{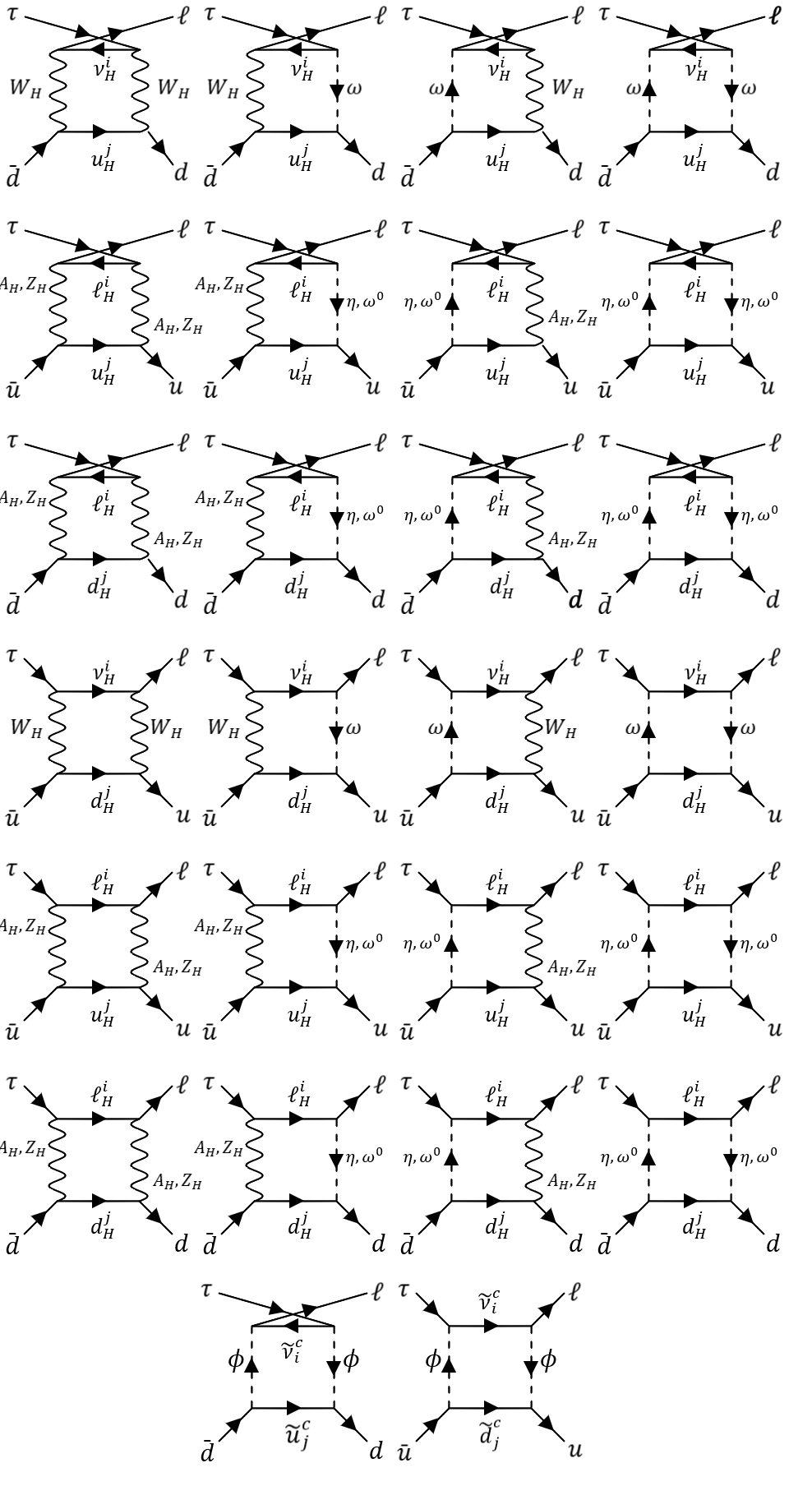}
    \caption{Box diagrams where T-odd particles,  partner leptons and quarks, are involved.}
    \label{fig:my_All_t_odds_diagrams}
\end{figure}

Only the box diagram involving  interaction between $\Phi^{+} \overline{\tilde{\nu}_{i}^{c}}\ell_{j}$ with $\Phi^{+} \overline{\tilde{u}_{i}^{c}}d_{j}$ (and its h.c.) contributes. We show in Figure \ref{fig:my_All_t_odds_diagrams} all box diagrams that appear when T-odd and partner fermions are considered. 

The box amplitude is defined as
\begin{equation}
    \mathcal{M}^{\mathrm{T-odd}}_{\mathrm{box}} = e^{2}B^{q}_{L}(0)\overline{\mu}(p')\gamma^{\mu}P_{L}\tau(p)\overline{q}(p_{q})\gamma_{\mu}P_{L}q(p_{\overline{q}}),
    \label{eq7.1.27}
\end{equation}
where $q = \{ u,d,s\}$. 

Therefore, in accordance to ref.~\cite{delAguila:2019htj}, the form factors from box diagrams are given by
{\small
\begin{align}
    B_{L}^{u} = & \frac{\alpha_{W}}{32\pi s_{W}^{2}} \left\lbrace \frac{1}{M_{W}^{2}} \frac{\upsilon^{2}}{4f^{2}} \sum_{i,j} \chi_{ij}^{u} \left[ - \left( 8 + \frac{1}{2} x_{i}^{\ell_{H}} y_{j}^{d_{H}} \right) \bar{d}_{0}(x_{i}^{\ell_{H}}, y_{j}^{d_{H}}) + 4 x_{i}^{\ell_{H}} y_{j}^{d_{H}} d_{0}(x_{i}^{\ell_{H}}, y_{j}^{d_{H}})  \right. \right. \nonumber \\
    & \left. \left. - \frac{3}{2} \bar{d}_{0} (x_{i}^{\nu_{H}}, y_{j}^{d_{H}}) - \frac{3}{50a} \bar{d}_{0} (ax_{i}^{\nu_{H}}, ay_{j}^{d_{H}}) + \frac{3}{5} \bar{d}_{0} (ax_{i}^{\nu_{H}}, ay_{j}^{d_{H}},a)  \right] + \frac{1}{f^{2}} \sum_{i,j} \bar{\chi}_{ij}^{u} \bar{d}_{0} ( \tilde{x}_{i}, \tilde{y}_{j}^{\tilde{d}^{c}} ) \right\rbrace, \nonumber \\
    B_{L}^{d} = & \frac{\alpha_{W}}{32\pi s_{W}^{2}} \left\lbrace \frac{1}{M_{W}^{2}} \frac{\upsilon^{2}}{4f^{2}} \sum_{i,j} \chi_{ij}^{d} \left[ \left( 2 + \frac{1}{2} x_{i}^{\ell_{H}} y_{j}^{u_{H}} \right) \bar{d}_{0}(x_{i}^{\ell_{H}}, y_{j}^{u_{H}}) - 4 x_{i}^{\ell_{H}} y_{j}^{u_{H}} d_{0}(x_{i}^{\ell_{H}}, y_{j}^{u_{H}})  \right. \right. \nonumber \\
    & \left. \left. - \frac{3}{2} \bar{d}_{0} (x_{i}^{\nu_{H}}, y_{j}^{u_{H}}) - \frac{3}{50a} \bar{d}_{0} (ax_{i}^{\nu_{H}}, ay_{j}^{u_{H}}) - \frac{3}{5} \bar{d}_{0} (ax_{i}^{\nu_{H}}, ay_{j}^{u_{H}},a)  \right] + \frac{5}{f^{2}} \sum_{i,j} \bar{\chi}_{ij}^{d} \bar{d}_{0} (\tilde{x}_{i}, \tilde{y}_{j}^{\tilde{u}^{c}} ) \right\rbrace,
    \label{eq7.1.28}
\end{align}
}
with $a = M_{W_{H}}^{2}/M_{A_{H}}^{2} = 5c_{W}^{2}/s^{2}_{W}$ and the four-point functions are expressed in Appendix \ref{App: A}. 

The mixing coefficients involve mirror lepton mixing matrices as well as mirror quark ones
\begin{equation}
    \chi_{ij}^{u} = V_{H\ell}^{i\mu\dagger}V_{H\ell}^{i\tau} V_{Hq}^{ju\dagger}V_{Hq}^{ju} , \quad \chi_{ij}^{d} = V_{H\ell}^{i\mu\dagger}V_{H\ell}^{i\tau} V_{Hq}^{jd\dagger}V_{Hq}^{jd},
    \label{eq7.1.29}
\end{equation}
and
\begin{align}
    \bar{\chi}_{ij}^{u} & = \sum_{k,n,r,s} V_{H\ell}^{k\mu\dagger} \frac{m_{\ell_{Hk}}}{M_{W_{H}}} W_{ki}^{\dagger} W_{in} \frac{m_{\ell_{Hn}}}{M_{W_{H}}} V_{H\ell}^{n\tau} V_{Hq}^{ru\dagger} \frac{m_{d_{Hr}}}{M_{W_{H}}} W_{rj}^{q \dagger} W_{js}^{q} \frac{m_{d_{Hs}}}{M_{W_{H}}} V_{Hq}^{su}, \nonumber \\
    \bar{\chi}_{ij}^{d} & = \sum_{k,n,r,s} V_{H\ell}^{k\mu\dagger} \frac{m_{\ell_{Hk}}}{M_{W_{H}}} W_{ki}^{\dagger} W_{in} \frac{m_{\ell_{Hn}}}{M_{W_{H}}} V_{H\ell}^{n\tau} V_{Hq}^{rd\dagger} \frac{m_{u_{Hr}}}{M_{W_{H}}} W_{rj}^{q \dagger} W_{js}^{q} \frac{m_{u_{Hs}}}{M_{W_{H}}} V_{Hq}^{sd}.
    \label{eq7.1.30}
\end{align}
In analogy to the lepton sector, the misalignment between the partner and the mirror quarks  mass eigenstates, as well as that  between the mirror and SM quarks are parametrized by the corresponding $3 \times 3$ unitary matrices, $V_{Hq}^{i}$ and $W_{ij}^{q}$, respectively.

\subsubsection{Majorana contribution}

Now, we are going to compute the $\mathcal{M}^{\mathrm{Maj}}$ contribution. As we showed in eq.(\ref{eq7.1.2}), it is composed by three parts 
\begin{equation}
    \mathcal{M}^{\mathrm{Maj}} = \mathcal{M}^{\mathrm{Maj}}_{\gamma} + \mathcal{M}^{\mathrm{Maj}}_{Z} + \mathcal{M}^{\mathrm{Maj}}_{\mathrm{box}},
    \label{eq7.1.41}
\end{equation}
$\gamma$ and $Z$ penguin diagrams and box diagrams which involve Majorana neutrinos. In this case we will work with the assumption that light Majorana neutrinos are massless, therefore just heavy Majorana neutrinos are taken into account.

The contributions from $\gamma-$ and $Z-$penguin diagrams and box diagrams are very similar to the ones for $\mu \to e$ conversion in nuclei:
\begin{align}
         \mathcal{M}^{\mathrm{Maj}}_{\gamma} & = \frac{e^{2}}{Q^{2}} \overline{\mu}(p') [i2F_{M}^{\gamma}(Q^{2})P_{R}\sigma^{\mu\nu}Q_{\nu} + F_{L}^{\gamma}(Q^{2})\gamma^{\mu}P_{L}]\tau(p) \overline{q}(p_{q}) \gamma_{\mu}Q_{q} q(p_{\bar{q}}), \nonumber \\
        \mathcal{M}_{Z}^{\mathrm{Maj}} & = \frac{e^{2}}{M_{Z}^{2}} \overline{\mu}(p')[\gamma^{\mu} (F_{L}^{Z}P_{L} + F_{R}^{Z}P_{R})]\tau(p)\overline{q}(p_{q})[\gamma_{\mu}\left(g_{Lq}^{Z}P_{L}+g_{Rq}^{Z}P_{R} \right)]q(p_{\bar{q}}), \nonumber \\
        \mathcal{M}^{\mathrm{Maj}}_{\mathrm{box}} & = e^{2}B_{L}^{q}(0) \overline{\mu}(p') \gamma^{\mu}P_{L}\tau(p)\overline{q}(p_{q}) \gamma_{\mu}P_{L}q(p_{\bar{q}}),
        \label{eq7.1.42}
\end{align}
with $Q_{q}$, the electric charge matrix, given by
\begin{equation}
    Q_{q} = \frac{1}{3} \left( \begin{array}{ccc}
        2 & &  \\
         & -1 & \\
          &  & -1
    \end{array} \right),
    \label{eq7.1.4}
\end{equation}
in units of $|e|$
and the couplings $g_{L(R)q}^{Z}$ read \cite{HernandezTome}
\begin{equation}
    \begin{split}
        g_{Lu}^{Z} & = \frac{1 - \frac{4}{3}s_{W}^{2}}{2s_{W}c_{W}}, \qquad g_{Ru}^{Z} = - \frac{2s_{W}}{3c_{W}}, \\
        g_{Ld}^{Z} & = \frac{-1 + \frac{2}{3}s_{W}^{2}}{2s_{W}c_{W}}, \qquad g_{Rd}^{Z} = \frac{s_{W}}{3c_{W}}.
    \end{split}
    \label{eq4.3.3}
\end{equation}

\begin{figure}[!ht]
    \centering
    \includegraphics[scale = 0.20]{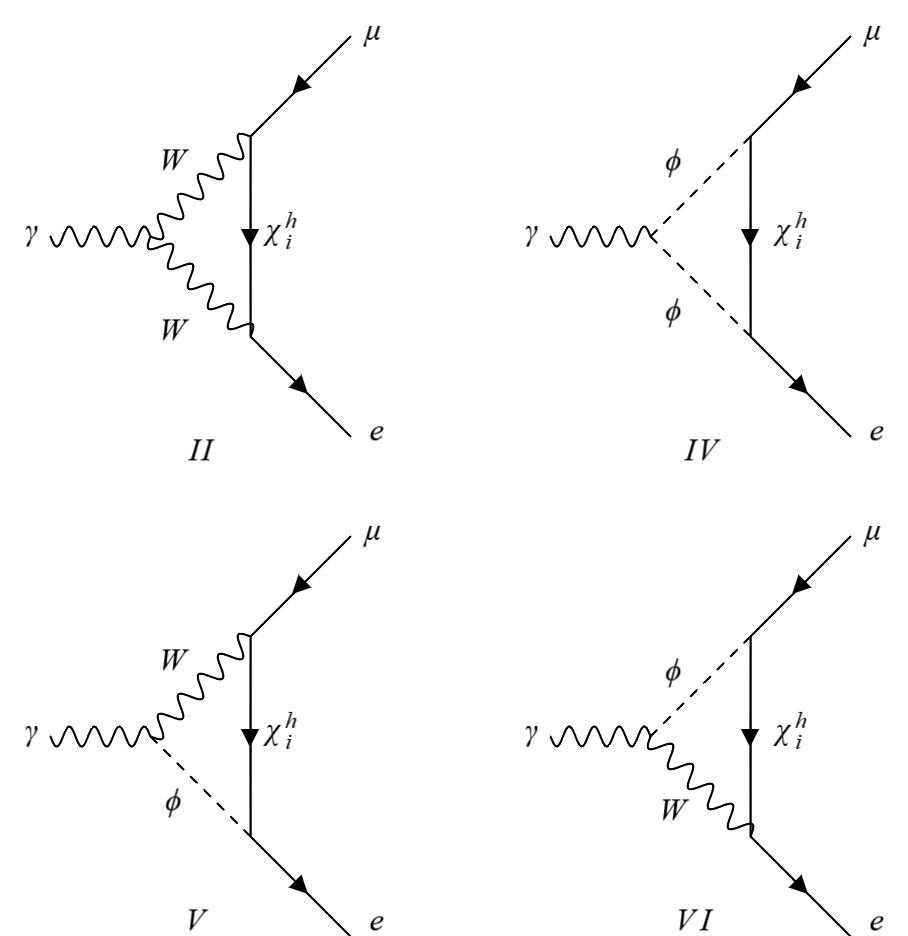}
    \caption{Feynman diagrams involved in the $\mu \rightarrow e\gamma$ decay considering heavy neutrinos. We have to take into account the self-energy diagrams, additionally. $\phi$ are the would-be Goldstones absorbed by $W$ boson in the unitary gauge.}
    \label{fig:33}
\end{figure}

\begin{figure}[!ht]
    \centering
    \includegraphics[scale = 0.39]{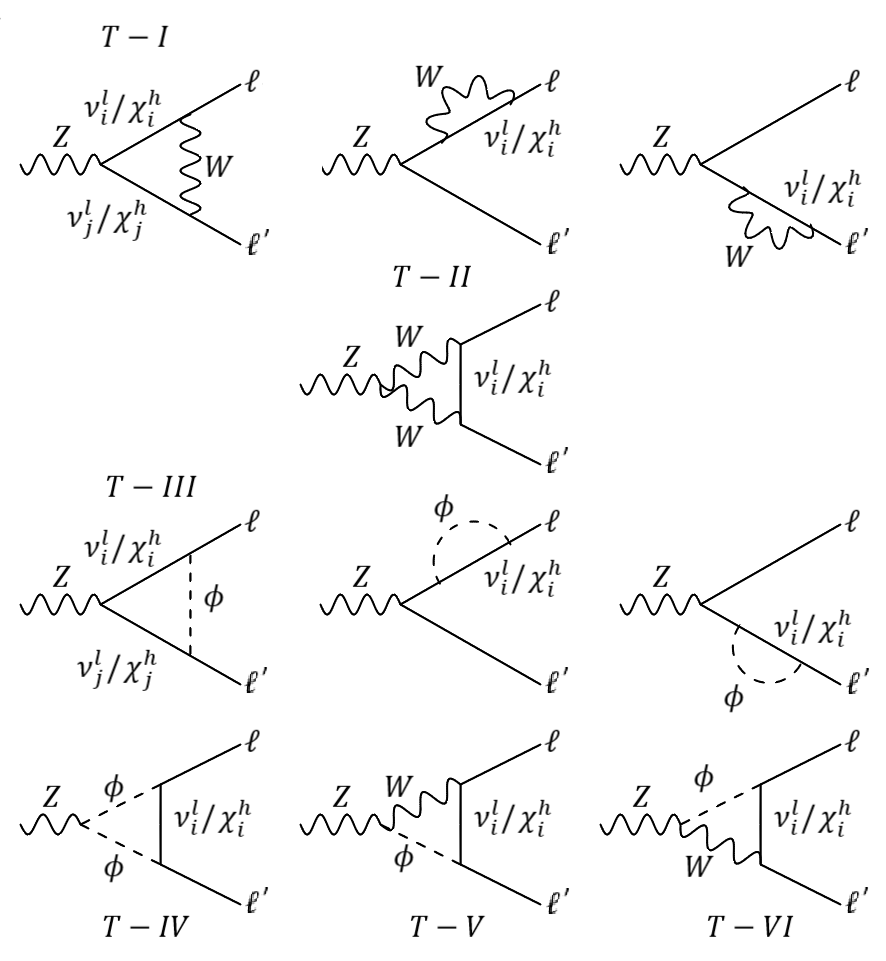}    \caption{Z penguins  diagrams that contribute to the decay. Diagrams $T-I$ and $T-III$ allow to mix light $(\nu^{l}_{i})$ and heavy Majorana neutrinos $(\chi^{h}_{i})$.}
    \label{fig:Z_penguins_Majorana}
\end{figure}

The form factors $F_{M}^{\gamma}$ (from Feynman diagrams II, IV, V and VI in  Figure \ref{fig:topologies}), $F_{L}^{\gamma}$ (from diagrams in Figure \ref{fig:33}) and $F_{L}^{Z}$ (see Figure \ref{fig:Z_penguins_Majorana}) are given by (omitting the light Majorana neutrinos contribution):
{\small
\begin{align}
    F_{M}^{\gamma} = F_{M}^{\chi^{h}} & = \frac{\alpha_{W}}{16\pi} \frac{m_{\tau}}{M_{W}^{2}} \sum_{j=1}^{3} \theta_{\mu j}\theta^{\dagger}_{\tau j} F_{M}^{\chi^{h}}(z_{j},Q^{2}), \nonumber \\
    F_{L}^{\gamma} = F_{L}^{\chi^{h}} & = \frac{\alpha_{W}}{8\pi } \sum_{j=1}^{3} \theta_{\mu j}\theta^{\dagger}_{\tau j}F_{L}^{\chi^{h}}(z_{j}, Q^{2}), \nonumber \\
    F_{L}^{Z-\chi^{h}}(Q^{2}) & =  \frac{\alpha_{W}}{8\pi c_{W}s_{W}} \sum_{i,j=1}^{3} \left[ \theta_{\mu i} \theta^{\dagger}_{\tau i} F^{h}(z_{i};Q^{2})  + \theta_{\mu j} S_{ji} \theta^{\dagger}_{\tau i}\left(G^{h}(z_{i},z_{j};Q^{2}) + \frac{1}{\sqrt{z_{i}z_{j}}} H^{h}(z_{i},z_{j};Q^{2}) \right) \right]\,.
    \label{eq7.1.43}
\end{align}
}
The functions which define the eq. (\ref{eq7.1.43}) are expressed explicitly in Appendix \ref{App: B}.

\begin{figure}[!ht]
    \centering
    \includegraphics[scale=0.35]{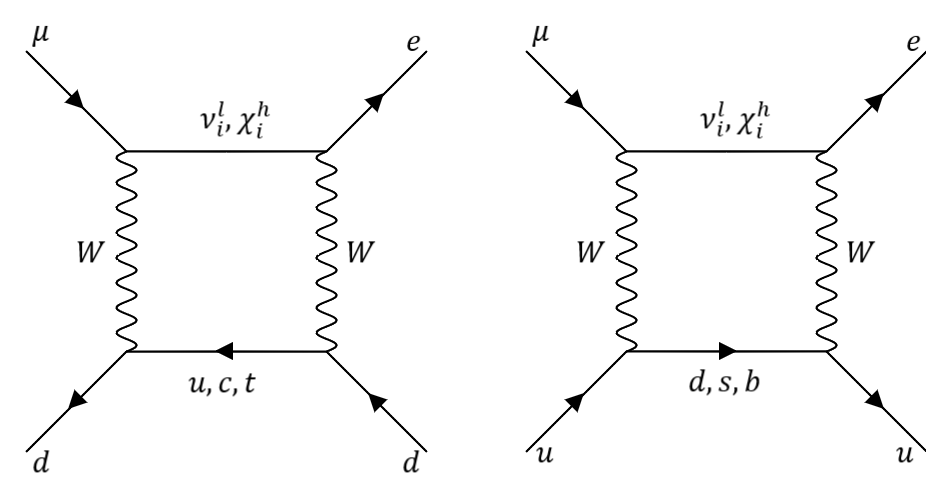}
    \caption{Box diagrams contributing to $\mu-e$ conversion in nuclei considering light-heavy Majorana neutrinos.}
    \label{fig:mue_conversion}
\end{figure}

For box diagrams we just consider the contribution coming from heavy Majorana neutrinos $\chi^{h}$ in Figure \ref{fig:mue_conversion}, which reads
\begin{equation}
    B_{L}^{d} = \frac{\alpha_{W}}{16 \pi M_{W}^{2} s_{W}^{2}} \sum_{i,j=1}^{3} \theta_{\mu i}\theta^{\dagger}_{\tau i} |V_{jd}|^{2}f_{B_{d}} \left(z_{i}, y_{j}^{u} \right),
    \label{eq7.1.45}
\end{equation}
\begin{equation}
    B_{L}^{u} = \frac{\alpha_{W}}{16 \pi M_{W}^{2} s_{W}^{2}} \sum_{i,j=1}^{3} \theta_{\mu i} \theta^{\dagger}_{\tau i} |V_{uj}|^{2}f_{B_{u}} (z_{i}, y_{j}^{d} ),
    \label{eq7.1.46}
\end{equation}
where $V_{ij}$ is the CKM matrix.
Neglecting all quark masses, except that of the top quark, and defining $y_{t} = m_{t}^{2}/M_{W}^{2}$, we may write \cite{HernandezTome}:
\begin{equation}
    \begin{split}
        \sum_{j=1}^{3}|V_{jd}|^{2}f_{B_{d}} \left( z_{i},y_{j}^{u} \right) = |V_{td}|^{2} \left[ f_{B_{d}}(z_{i},y_{t}) - f_{B_{d}}(z_{i},0)  \right] + f_{B_{d}}(z_{i},0),
    \end{split}
    \label{eq7.1.48.1}
\end{equation}
\begin{equation}
    \begin{split}
        \sum_{i=j}^{3}|V_{uj}|^{2}f_{B_{u}}( z_{i},y_{j}^{d}) = f_{B_{u}}(z_{i},0).
    \end{split}
    \label{eq7.1.48.2}
\end{equation}

\subsection{Hadronization}
\label{sec:Had}

Tau decays we are considering have as final states pseudoscalar mesons and vector resonances. Hadronization of quark bilinears gives  rise to the final-state hadrons.

Resonance Chiral Theory (R$\chi$T) \cite{Ecker:1988te, Ecker:1989yg}, that naturally includes Chiral Perturbation Theory ($\chi$PT) \cite{Gasser:1983yg, Gasser:1984gg}, considers the resonances as active degrees of freedom into the Lagrangian since they drive the dynamics of the processes. We are working under R$\chi$T scheme in order to hadronize the relevant currents involved in our analysis. For more details on the procedure followed in this part, see refs. \cite{Arganda:2008jj,Husek:2020fru} where the definitions of all expressions are fully given. In Appendix \ref{app:C} we write all the useful tools for the development which is shown next.

We remind that the complete amplitude has two contributions:
\begin{equation}
    \mathcal{M} = \mathcal{M}^{\mathrm{T-odd}} + \mathcal{M}^{\mathrm{Maj}},
    \label{eq7.1.49}
\end{equation}
where each one receives contributions coming from $\gamma-$ and $Z-$penguins, and box diagrams.\\
The $F_{M}^{\gamma}$ form factor, considering both contributions of T-odd leptons (eq.~(\ref{eq7.1.8})) and Majorana neutrinos (eq.~(\ref{eq7.1.43})), is written as follows 

\begin{align}
    F_{M}^{\gamma} = & \frac{\alpha_{W}}{16\pi}\frac{m_{\tau}}{M_{W}^{2}} \left\lbrace \sum_{i=1}^{3} \left(\theta_{\mu i}\theta^{\dagger}_{\tau i}F_{M}^{\chi^{h}}(z_{i},Q^{2}) \right. \right. \nonumber \\
    & \left. \left. + \frac{\upsilon^{2}}{4f^{2}} V_{H\ell}^{i\mu*}V_{H\ell}^{i\tau} \left[ F_{W_{H}}(x_{i}^{\ell_{H}},Q^{2}) + F_{Z_{H}}(x_{i}^{\nu_{H}},Q^{2}) + \frac{1}{5}F_{Z_{H}}(ax_{i}^{\nu_{H}},Q^{2}) \right] \right) \right. \nonumber \\
    & \left. + \frac{1}{M_{h}^{2}} \frac{\upsilon^{4}}{4f^{2}} \sum_{i,j,k = 1}^{3} V_{H \ell}^{i\mu \dagger}\kappa_{ii}W_{ij}^{\dagger}W_{jk}\kappa_{kk}V_{H\ell}^{k\tau} \left[F_{\tilde{\nu}^{c}}\left( \tilde{x}_{j},Q^{2} \right)+F_{\tilde{\ell}^{c}}\left(\tilde{x}_{j},Q^{2} \right)\right] \right\rbrace\,,
    \label{eq7.1.50}
\end{align}

Similarly, $F_{L}^{\gamma}$ can be written from eqs. (\ref{eq7.1.11}) and (\ref{eq7.1.43}) as
\begin{align}
    F_{L}^{\gamma} = & \frac{\alpha_{W}}{4\pi M_{W}^{2}} \left\lbrace \sum_{i=1}^{3} \left( \frac{1}{2} \theta_{\mu i}\theta^{\dagger}_{\tau i}F_{L}^{\chi^{h}}(z_{i},Q^{2}) \right. \right. \nonumber \\
    & \left. \left.+ M_{W}^{2} \frac{Q^{2}}{M_{W_{H}}^{2}} V_{H\ell}^{i\mu*}V_{H\ell}^{i\tau} \left[G^{(1)}_{W}(x_{i}^{\ell_{H}}) + G^{(1)}_{Z}(x_{i}^{\nu_{H}}) + \frac{1}{5}G_{Z}^{(1)}(ax_{i}^{\nu_{H}}) \right] \right) \right. \nonumber \\
    & \left. + \upsilon^{2} \frac{M_{W}^{2}}{M_{h}^{2}} \frac{Q^{2}}{M_{W_{H}}^{2}} \sum_{i,j,k=1}^{3} V_{H \ell}^{i\mu \dagger}\kappa_{ii}W_{ij}^{\dagger}W_{jk}\kappa_{kk}V_{H\ell}^{k\tau}\left[ G_{\tilde{\nu}^{c}}^{(1)}\left(\tilde{x}_{j}\right) + G_{\tilde{\ell}^{c}}^{(1)}\left(\tilde{x}_{j}\right) \right]\right\rbrace.
    \label{eq7.1.51}
\end{align}
The form factor coming from $Z-$penguin diagrams, taking into account T-odd particles (eq. (\ref{eq7.1.16})) as well as Majorana neutrinos (eq. (\ref{eq7.1.43})) reads
{\small
\begin{align}
    F_{L}^{Z} = & \frac{\alpha_{W}}{8\pi c_{W}s_{W}} \sum_{i,j=1}^{3} \left[ \theta_{\mu i} \theta^{\dagger}_{\tau i} F^{h}(z_{i};Q^{2})  + \theta_{\mu j} S_{ji} \theta^{\dagger}_{\tau i}\left(G^{h}(z_{i},z_{j};Q^{2}) + \frac{1}{\sqrt{z_{i}z_{j}}} H^{h}(z_{i},z_{j};Q^{2}) \right) \right. \nonumber \\
    & \left. + \sum_{i}V_{H\ell}^{i\mu\dagger}V_{H\ell}^{i\tau} \left\lbrace \frac{\upsilon^{2}}{8f^{2}} H_{L}^{W(0)}(x_{i}^{\ell_{H}}) + \frac{Q^{2}}{M_{W_{H}}^{2}} \left[ H_{L}^{W}(x_{i}^{\ell_{H}}) + (1-2c_{W}^{2}) \left( \frac{1}{5} H_{L}^{A/Z}(ax_{i}^{\nu_{H}}) + H_{L}^{A/Z}(x_{i}^{\nu_{H}}) \right) \right] \right\rbrace \right. \nonumber \\
    & \left. + \frac{Q^{2}}{M_{\Phi}^{2}} \frac{\upsilon^{2}}{2M_{W}^{2}}\sum_{ijk} V_{H \ell}^{i\mu \dagger}\kappa_{ii}W_{ij}^{\dagger}W_{jk}\kappa_{kk}V_{H\ell}^{k\tau} \left[ H_{L}^{\bar{\nu}}\left(\tilde{x}_{j}\right) + (1-2c_{W}^{2})H_{L}^{\bar{\ell}}\left(\tilde{x}_{j}\right) \right] \right].
    \label{eq7.1.52}
\end{align}
}
For box diagrams the $B_{L}^{q}$ form factors are given by eqs. (\ref{eq7.1.28}), (\ref{eq7.1.45}) and (\ref{eq7.1.46}), yielding
{\small
\begin{align}
    B_{L}^{u} & = \frac{\alpha_{W}}{16\pi M_{W}^{2} s_{W}^{2}} \left\lbrace \sum_{i,j=1}^{3}  \theta_{\mu i} \theta^{\dagger}_{\tau i} f_{B_{u}}(z_{i},0)  \right. \nonumber \\ 
    & \left. - \frac{\upsilon^{2}}{8f^{2}} \sum_{i,j} \chi_{ij}^{u} \left[ \left( 8 + \frac{1}{2} x_{i}^{\ell_{H}} y_{j}^{d_{H}} \right) \bar{d}_{0}(x_{i}^{\ell_{H}}, y_{j}^{d_{H}}) - 4 x_{i}^{\ell_{H}} y_{j}^{d_{H}} d_{0}(x_{i}^{\ell_{H}}, y_{j}^{d_{H}})  \right. \right. \nonumber \\
    & \left. \left. + \frac{3}{2} \bar{d}_{0} (x_{i}^{\nu_{H}} y_{j}^{d_{H}}) + \frac{3}{50a} \bar{d}_{0} (ax_{i}^{\nu_{H}}, ay_{j}^{d_{H}}) - \frac{3}{5} \bar{d}_{0} (ax_{i}^{\nu_{H}}, ay_{j}^{d_{H}},a)  \right] + \frac{M_{W}^{2}}{2f^{2}} \sum_{i,j} \bar{\chi}_{ij}^{u} \bar{d}_{0} \left( \tilde{x}_{i},\tilde{y}_{j}^{\tilde{d}^{c}} \right) \right\rbrace\,,
    \label{eq7.1.53}
\end{align}

\begin{align}
    B_{L}^{d} & = \frac{\alpha_{W}}{16 \pi M_{W}^{2} s_{W}^{2}} \left\lbrace \sum_{i=1}^{3} \theta_{\mu i}\theta^{\dagger}_{\tau i} \left( |V_{td}|^{2} \left[ f_{B_{d}}(z_{i},y_{t}) - f_{B_{d}}(z_{i},0)  \right] + f_{B_{d}}(z_{i},0) \right) \right. \nonumber \\
    & \left. + \frac{\upsilon^{2}}{8f^{2}} \sum_{i,j} \chi_{ij}^{d} \left[ \left( 2 + \frac{1}{2} x_{i}^{\ell_{H}} y_{j}^{u_{H}} \right) \bar{d}_{0}(x_{i}^{\ell_{H}}, y_{j}^{u_{H}}) - 4 x_{i}^{\ell_{H}} y_{j}^{u_{H}} d_{0}(x_{i}^{\ell_{H}}, y_{j}^{u_{H}})  \right. \right. \nonumber \\
    & \left. \left. - \frac{3}{2} \bar{d}_{0} (x_{i}^{\nu_{H}}, y_{j}^{u_{H}}) - \frac{3}{50a} \bar{d}_{0} (ax_{i}^{\nu_{H}}, ay_{j}^{u_{H}}) - \frac{3}{5} \bar{d}_{0} (ax_{i}^{\nu_{H}}, ay_{j}^{u_{H}},a)  \right] + \frac{5}{2} \frac{M_{W}^{2}}{f^{2}} \sum_{i,j} \bar{\chi}_{ij}^{d} \bar{d}_{0} \left( \tilde{x}_{i},\tilde{y}_{j}^{\tilde{u}^{c}} \right) \right\rbrace\,.
    \label{eq7.1.54}
\end{align}
}

The construction of the branching ratios for any decay mode $\tau \to \ell P$, $\tau \to \ell PP $, and $\tau \to \ell V$ can be developed with the expressions defined in Appendix \ref{app:C}.

\section{Phenomenology}
\label{sec:Pheno}

First of all, we recall the masses of particles which come from LHT that are involved in the processes under study \cite{delAguila:2008zu,Monika Blanke-066,FranciscoLluis}:
\begin{equation}
\begin{split}
        M_{W} & = \frac{e\upsilon}{2s_{W}} \left( 1 - \frac{\upsilon^{2}}{12f^{2}} \right), \quad M_{Z} = M_{W}/c_{W}, \quad \upsilon \simeq 246 \ \mathrm{GeV}, \quad (\rho \ \mathrm{factor \ is \ conserved}),\\
        M_{W_{H}} & = M_{Z_{H}} = \frac{ef}{s_{W}} \left( 1 -\frac{\upsilon^{2}}{8f^{2}} \right), \quad M_{A_{H}} = \frac{ef}{\sqrt{5}c_{W}} \left( 1 - \frac{5\upsilon^{2}}{8f^{2}} \right), \quad M_{\Phi} = \sqrt{2}M_{h}\frac{f}{\upsilon},\\
        m_{\ell_{H}^{i}} & = \sqrt{2}\kappa_{ii}f \equiv m_{Hi}, \quad m_{\nu^{i}_{H}} = m_{Hi} \left( 1 - \frac{\upsilon^{2}}{8f^{2}} \right), \quad m_{\ell^{c}, \nu^{c}} = \kappa_{2},
\end{split} 
\label{eq7.6.1}
\end{equation}
with $M_{h}$ being the mass of the SM Higgs scalar, $\kappa_{ii}$ the diagonal entries of the $\kappa$ matrix (see eq. (\ref{eq7.1.01})) (similarly for the masses of T-odd quarks with $\kappa_{ii}^{q}$ instead of $\kappa_{ii}$ and replacing $d_{H}$-quark by $\ell_{H}$ and $u_{H}$-quark by $\nu_{H}$) and $\kappa_{2}$ the mass matrix of partner leptons from eq. (\ref{eq7.1.01}) ($\kappa_{2}^{q}$ similarly for partner quarks of u and d types). Due to $M_{h} = 125.25 \pm 0.17 \ \mathrm{GeV}$ \cite{PDG} and $\upsilon \simeq 246 \ \mathrm{GeV}$, we can approximate
\begin{equation}
    M_{\Phi} = \sqrt{2}M_{h}\frac{f}{\upsilon} \approx \frac{\sqrt{2}}{2} f\,.
    \label{eq7.6.2}
\end{equation}
So far, the free parameters are: $f$ (scale of new physics); $\kappa_{ii}$ and $\kappa_{ii}^{q}$ (giving masses for T-odd leptons and quarks); $\kappa_{2}$ and $\kappa_{2}^{q}$, mass matrices for partner leptons.\\
    The expressions $\chi_{ij}^{u}$, $\chi_{ij}^{d}$, $\bar{\chi}_{ij}^{u}$ and $\bar{\chi}_{ij}^{d}$ (see eq. (\ref{eq7.1.30})) describe the interaction vertices from box diagrams, those can be re-written in terms of free parameters as follows 
\begin{align}
    \bar{\chi}_{ij}^{u} & = \frac{\upsilon^{4}}{4M_{W}^{4}} \sum_{k,n,r,s} V_{H\ell}^{k\mu\dagger} \kappa_{kk} W_{ki}^{\dagger} W_{in} \kappa_{nn} V_{H\ell}^{n\tau} V_{Hq}^{ru\dagger} \kappa^{d}_{rr} W_{rj}^{q \dagger} W_{js}^{q} \kappa^{d}_{ss} V_{Hq}^{su}, \nonumber \\
    \bar{\chi}_{ij}^{d} & = \frac{\upsilon^{4}}{4M_{W}^{4}} \left( 1 - \frac{\upsilon^{2}}{8f^{2}} \right)^{2} \sum_{k,n,r,s} V_{H\ell}^{k\mu\dagger} \kappa_{kk} W_{ki}^{\dagger} W_{in} \kappa_{nn} V_{H\ell}^{n\tau} V_{Hq}^{rd\dagger} \kappa^{u}_{rr} W_{rj}^{q \dagger} W_{js}^{q} \kappa^{u}_{ss} V_{Hq}^{sd}\,,
    \label{eq7.6.7}
\end{align}
where we see a small shift between interaction vertices of order $\mathcal{O}(\upsilon^{2}/8f^{2})$. The mixing matrices of heavy Majorana neutrinos are bounded by \cite{Pacheco}
\begin{equation}
    |\theta_{e j}\theta^{\dagger}_{\tau j}| < 0.95 \times 10^{-2},\quad
  |\theta_{\mu j}\theta^{\dagger}_{\tau j}| < 0.011.
    \label{eq7.6.8}
\end{equation}
Considering just mixing between two lepton families for simplicity, the mixing matrix of T-odd leptons $(V_{H\ell}^{i\mu*}V_{H\ell}^{i\tau})$ and the mixing matrix among partner leptons $(W_{ij}^{i\mu\dagger}W_{jk}^{i\tau})$ can be parameterized as follows \cite{delAguila:2019htj}
\begin{equation}
    V = \left( \begin{array}{ccc}
        1 & 0 & 0 \\
        0 & \cos\theta_{V} & \sin \theta_{V} \\
        0 & - \sin \theta_{V} & \cos \theta_{V}
    \end{array} \right) \ , \quad W = \left( \begin{array}{ccc}
        1 & 0 & 0 \\
        0 & \cos\theta_{W} & \sin \theta_{W} \\
        0 & - \sin \theta_{W} & \cos \theta_{W}
    \end{array} \right),
    \label{eq7.6.9}
\end{equation}
where $\theta_{V}, \theta_{W} \in [0, \pi / 2 ) $ is the physical range for the mixing angles and $\theta_{W}$ must not be confused with the weak-mixing ('Weinberg') angle. Before, we assumed $\mu - \tau$ mixing, and proceed similarly for the evaluation of processes with $\tau - e$ transitions 
(analogous mixings, in the top left $2\times2$ submatrix, can be used for quark contributions to $\mu\to e$ conversion in nuclei).

We will assume no extra quark mixing and degenerate heavy quarks, then $V_{H\ell}^{q}$ and $W^{q}$ will be the identity. Therefore, the other free parameter are: $\theta_{V}$, $\theta_{W}$ and neutral couplings of heavy Majorana neutrinos: $(\theta S \theta^{\dagger})_{\mu \tau}$.

For the form factors we do a consistent expansion on the squared transfer momenta over  the squared masses of heavy particles, $Q^{2}/\lambda^{2}$, being $\lambda = \{ M_{W_{H}}, M_{Z_{H}}, M_{A_{H}}, M_{\Phi}, M_{i} \}$. This amounts to an expansion, at the largest, in the $m_{\tau}^{2}/f^{2}$ ratio.

For completeness and for the interest of the first one on its own, we include two analyses: first we do not assume heavy Majorana neutrinos contributions (this, within the LHT, was not considered before in the literature) and the second case adds the presence of these neutrinos arising from the Inverse See Saw mechanism, as seen in previous sections.

All Monte Carlo simulations for each process are intended to yield  mean values for the model free parameters and thereby give mean values for the branching ratios  that respect the current bounds reported in the PDG \cite{PDG}. These results should be representative of the possible model phenomenology under the assumptions that we have taken.

\subsection{Without Majorana Neutrinos Contribution}
\label{sec: Without Majorana Neutrinos Contribution}

We begin the discussion of our  results with the case without  Majorana neutrinos. The $\tau \to \ell P$ processes are computed in a single Monte Carlo simulation which runs them simultaneously. Because the structure of the branching ratios for the processes $\tau \to \ell PP$ and $\tau \to \ell V \ (\ell = e, \mu)$ is very similar, they have been computed jointly by another  single Monte Carlo simulation. We scan the model parameter space with the help of the SpaceMath package \cite{Arroyo-Urena:2020qup}, ensuring that all processes respect their current upper limits. 
The resulting mean values obtained from our analyses are shown in Tables \ref{tau_ellP_sin_Majoranas} and \ref{tau_ellPP_sin_Majorana} and discussed below.

\begin{table}[!ht]
    \scriptsize
    \centering
    \begin{tabular}{c|c|c|c}
    \hline \hline
    \multicolumn{4}{c}{\textbf{$\tau \to \ell P \ (\ell = e,\mu) \ (\mathrm{C.L.} = 90\%)$} without Majorana neutrinos contribution.} \\
    \hline \hline
    \multicolumn{2}{c|}{New physics (NP) scale (TeV)} & \multicolumn{2}{c}{Mixing angles} \\
    \hline
     $f$ & 1.49
     & $\theta_{V}$ & $42.78^{\circ}$ \\
     \cline{1-2}
    \multicolumn{2}{c|}{Branching ratio} & $\theta_{W}$ & $42.69^{\circ}$ \\
    \hline
     Br($\tau \to e \pi^{0}$) & $5.24
     \times 10^{-9}$  & \multicolumn{2}{|c}{Masses of partner leptons $(m_{\nu^{c}} = m_{\ell^{c}})$(TeV)} \\
     \cline{3-4}
     Br($\tau \to \mu \pi^{0}$) & $3.42
     \times 10^{-9}$ & $m_{\nu^{c}_{1}}$ & 3.12
     \\
     Br($\tau \to e \eta$) & $2.32
     \times 10^{-9}$ & $m_{\nu^{c}_{2}}$ & 3.15
     \\
     Br($\tau \to \mu \eta$) & $1.91
     \times 10^{-9}$ & $m_{\nu^{c}_{3}}$ & 3.37
     \\
     \cline{3-4}
     Br($\tau \to e \eta'$) & $2.20
     \times 10^{-8}$ & \multicolumn{2}{|c}{Masses of partner quarks $(m_{u^{c}} = m_{d^{c}})$ (TeV) } \\
     \cline{3-4}
     Br($\tau \to \mu \eta'$) & $1.79
     \times 10^{-8}$ & $m_{u^{c}_{i}}$ & 3.55
     \\
     \hline
     \multicolumn{2}{c|}{Masses of T-odd leptons (TeV)}&&\\
     \cline{1-2}
     $m_{\ell^{1}_{H}}$ & 2.11
     &  & 
     \\
     $m_{\ell^{2}_{H}}$ & 2.11
     &  & 
     \\
     $m_{\ell^{3}_{H}}$ & 2.12
     &  & 
     \\
     $m_{\nu^{1}_{H}}$ & 2.10
     & &  \\
     $m_{\nu^{2}_{H}}$ & 2.11
     &  & \\
     $m_{\nu^{3}_{H}}$ & 2.11
     &  &  \\
     \cline{1-2}
     \multicolumn{2}{c|}{Masses of T-odd quarks (TeV)}&  & \\
     \cline{1-2}
     $m_{d^{i}_{H}}$ & 2.71
     &  & \\
     $m_{u^{i}_{H}}$ & 2.70
     &  & \\
     \hline
    \end{tabular}
    \caption{Mean values for branching ratios, masses of
    LHT heavy particles, and mixing angles obtained by Monte Carlo simulation of $\tau \to \ell P \ (\ell = e,\mu)$ processes where Majorana neutrinos contribution is not considered.}
    \label{tau_ellP_sin_Majoranas}
\end{table}

\begin{table}[!ht]
    \scriptsize
    \centering
    \begin{tabular}{c|c|c|c}
    \hline \hline
    \multicolumn{4}{c}{\textbf{$\tau \to \ell PP, \ \ell V \ (\ell = e,\mu) \ (\mathrm{C.L.} = 90\%)$} without Majorana neutrinos contribution} \\
    \hline \hline
    \multicolumn{2}{c|}{New physics (NP) scale (TeV)} & \multicolumn{2}{c}{Mixing angles} \\
    \hline
     $f$ & 1.50
     & $\theta_{V}$ & $43.36^{\circ}$ \\
     \cline{1-2}
    \multicolumn{2}{c|}{Branching ratio} & $\theta_{W}$ & $41.50^{\circ}$ \\
    \hline
     Br($\tau \to e \pi^{+}\pi^{-}$) & $3.92
     \times 10^{-9}$  & \multicolumn{2}{|c}{Masses of partner leptons $(m_{\nu^{c}} = m_{\ell^{c}})$(TeV)} \\
     \cline{3-4}
     Br($\tau \to \mu \pi^{+}\pi^{-}$) & $3.96
     \times 10^{-9}$ & $m_{\nu^{c}_{1}}$ & 3.20
     \\
     Br($\tau \to e K^{+}K^{-}$) & $2.38
     \times 10^{-9}$ & $m_{\nu^{c}_{2}}$ & 3.15
     \\
     Br($\tau \to \mu K^{+}K^{-}$) & $2.85
     \times 10^{-9}$ & $m_{\nu^{c}_{3}}$ & 3.31
     \\
     \cline{3-4}
     Br($\tau \to e K^{0} \overline{K^{0}}$) & $1.15
     \times 10^{-9}$ & \multicolumn{2}{|c}{Masses of partner quarks $(m_{u^{c}} = m_{d^{c}})$ (TeV) } \\
     \cline{3-4}
     Br($\tau \to \mu K^{0} \overline{K^{0}}$) & $1.33
     \times 10^{-9}$ & $m_{u^{c}_{i}}$ & 3.32 \\
     Br($\tau \to e \rho$) & $1.10
     \times 10^{-9}$ & & \\
     Br($\tau \to \mu \rho$) & $1.12
     \times 10^{-9}$ & & \\
     Br($\tau \to e \phi$) & $1.77
     \times 10^{-9}$ & & \\
     Br($\tau \to \mu \phi$) & $1.87
     \times 10^{-9}$ & & \\
     \cline{1-2}
     \multicolumn{2}{c|}{Masses of T-odd leptons (TeV)}& &\\
     \cline{1-2}
     $m_{\ell^{1}_{H}}$ & 3.13
     &  & 
     \\
     $m_{\ell^{2}_{H}}$ & 2.99
     &  & 
     \\
     $m_{\ell^{3}_{H}}$ & 3.10
     &  & 
     \\
     $m_{\nu^{1}_{H}}$ & 3.12
     & & \\
     $m_{\nu^{2}_{H}}$ & 2.98
     &  & \\
     $m_{\nu^{3}_{H}}$ & 3.09
     &  &  \\
     \cline{1-2}
     \multicolumn{2}{c|}{Masses of T-odd quarks (TeV)}&  & \\
     \cline{1-2}
     $m_{d^{i}_{H}}$ & 2.92
     &  & \\
     $m_{u^{i}_{H}}$ & 2.91
     &  & \\
     \hline
    \end{tabular}
    \caption{Mean values for branching ratios, masses of
    LHT heavy particles, mixing angles and neutral couplings obtained by Monte Carlo simulation of $\tau \to \ell PP, \ \ell V \ (\ell = e,\mu)$ processes without Majorana neutrinos.}
    \label{tau_ellPP_sin_Majorana}
\end{table}


The NP scale coincides at the one percent level in both analyses, $f \sim 1.50$ TeV, supporting their individual consistency.

From Tables \ref{tau_ellP_sin_Majoranas} and \ref{tau_ellPP_sin_Majorana}, the mean values for the T-odd lepton masses in the $\tau \to \ell P$ modes are lighter than the $\tau \to \ell PP, \ \ell V $ ones by $\sim 1$ TeV. 
For $\tau \to \ell P$ processes T-odd leptons are lighter than T-odd quarks by $0.6$ TeV which means that $|\kappa_{ii}| < |\kappa_{ii}^{q}|$, 
(namely $|\kappa_{ii}| < 1.002 \ \mathrm{and} \ |\kappa_{ii}^{q}| < 1.282$). We recall that in our model we are considering degenerate  T-odd quarks for simplicity.

Unlike processes in Table \ref{tau_ellP_sin_Majoranas}, in the $\tau \to \ell PP, \ \ell V$ ones the T-odd leptons are heavier than the T-odd quarks. In both cases the T-odd quarks keep masses below 3 TeV, while for $\tau \to \ell PP, \ \ell V$ processes the mean values for T-odd lepton  masses exceed 3 TeV.

Regarding partner leptons $(m_{\nu_{i}^{c}})$ and partner quarks $(m_{u_{i}^{c}})$,  their masses are above $3$ TeV, being the latter the heaviest particles coming from LHT. We see that the mean  masses of partner leptons do not have a sizeable difference between $\tau \to \ell P$ and $\tau \to \ell PP, \ \ell V$ decay modes. The mean values of the partner quark masses differ by $\sim 6.5 \%$ between these tables (\ref{tau_ellP_sin_Majoranas} and \ref{tau_ellPP_sin_Majorana}).   

The mean values for the mixing angles $\theta_{V}$ and $\theta_{W}$ considering the results of both Tables \ref{tau_ellP_sin_Majoranas} and \ref{tau_ellPP_sin_Majorana} are  
$\theta_{V} \approx \theta_{W} \approx \pi/4.18$ 
. This result is close to maximize the LFV effects, since this happens when $\theta_{V} = \theta_{W} = \pi/4$.

\begin{figure}[!ht]
\centering
    \begin{minipage}{0.73\linewidth}
    \centering
     \includegraphics[width=\linewidth]{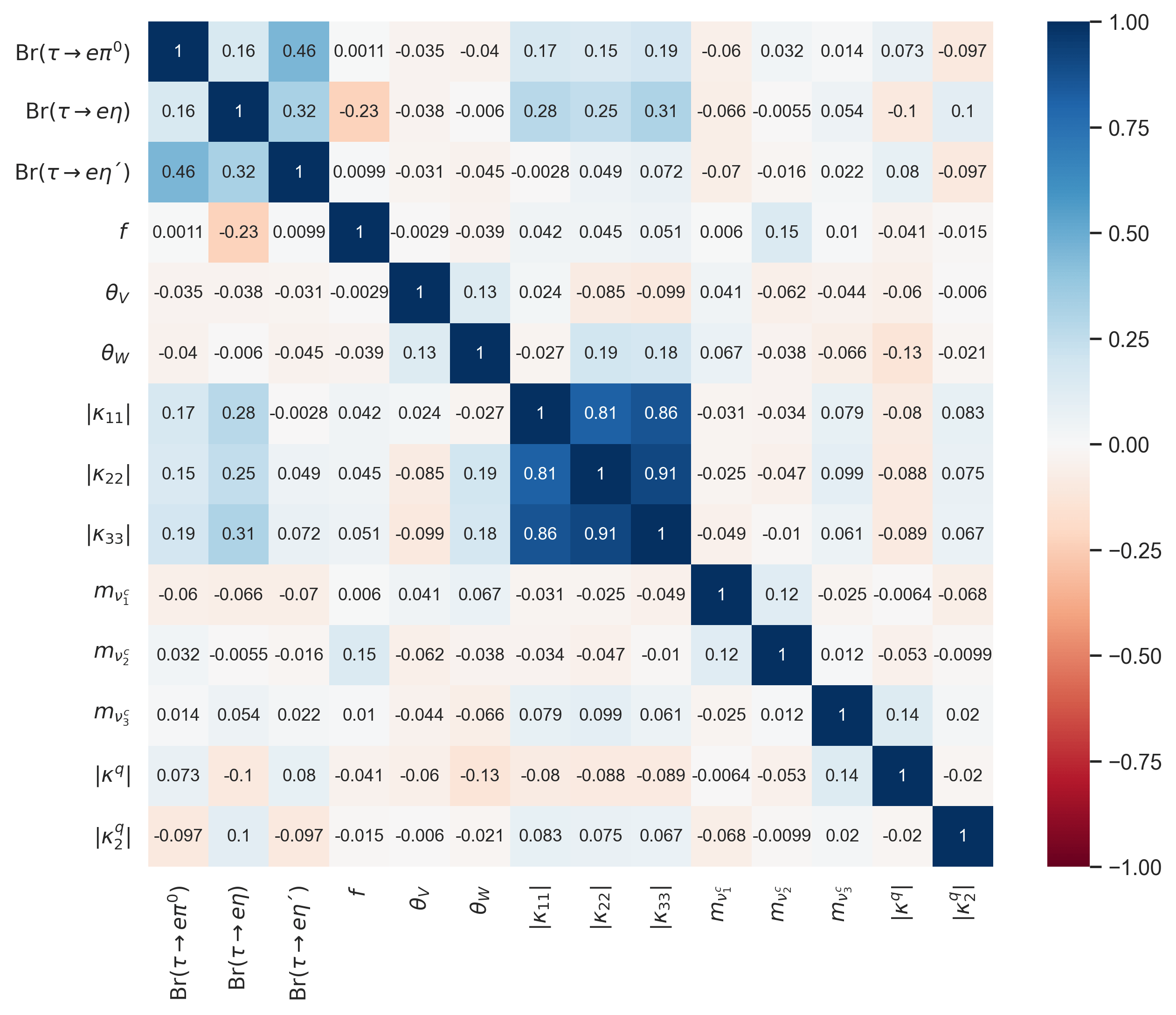}
     \caption{Heat map showing little  correlation among $\tau \to e P$ decays and their free parameters (case without Majorana neutrinos).}
     \label{fig:tau_eP_sin_Majoranas}
    \end{minipage}
    \hspace{1cm}
    \begin{minipage}{0.73\linewidth}
    \centering
    \includegraphics[width=\linewidth]{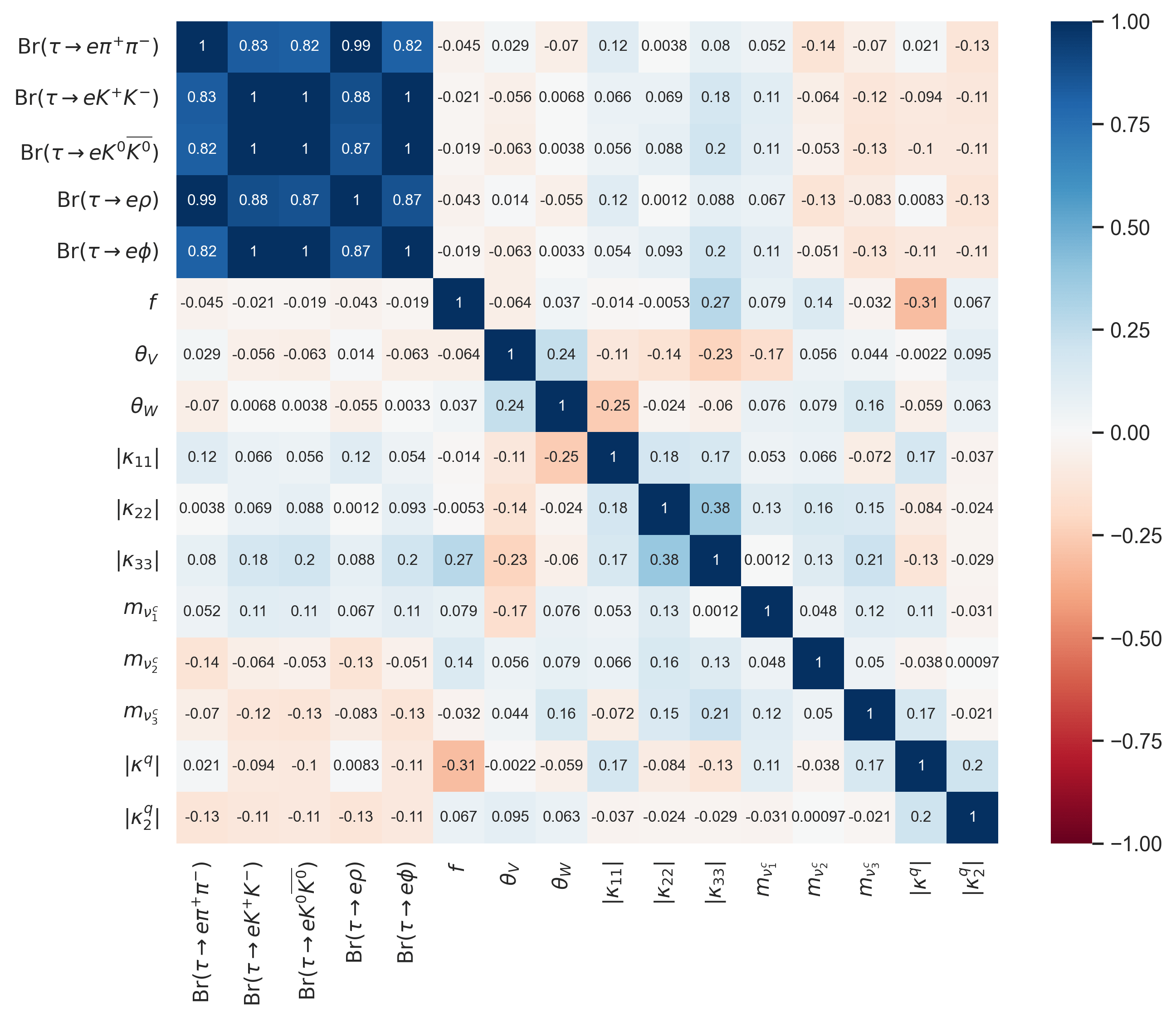}
     \caption{Heat map for $\tau \to e P P, \ e V$ decays and their free parameters not considering Majorana neutrinos.}
     \label{fig:tau_ePP_V_sin_Majorana}
    \end{minipage}
\end{figure}

In Figures \ref{fig:tau_eP_sin_Majoranas} and \ref{fig:tau_ePP_V_sin_Majorana} the correlations among branching ratios and their free parameters are shown for decay mode with $\ell = e$ (for $\ell = \mu$ the behavior is very similar and not shown). We notice that no branching ratio  (both heat maps) has a sizeable correlation with their free parameters. But in $\tau \to \ell P$ decay modes the correlations among the Yukawa couplings of T-odd leptons are high, which results in their masses being very similar, whereas that correlation vanishes in $\tau \to \ell PP, \ \ell V$. We  highlight that all branching fractions are highly correlated for any type of decay. 

\begin{figure}[!ht]
\centering
    \begin{minipage}{0.45\linewidth}
    \centering
        \includegraphics[width=\linewidth]{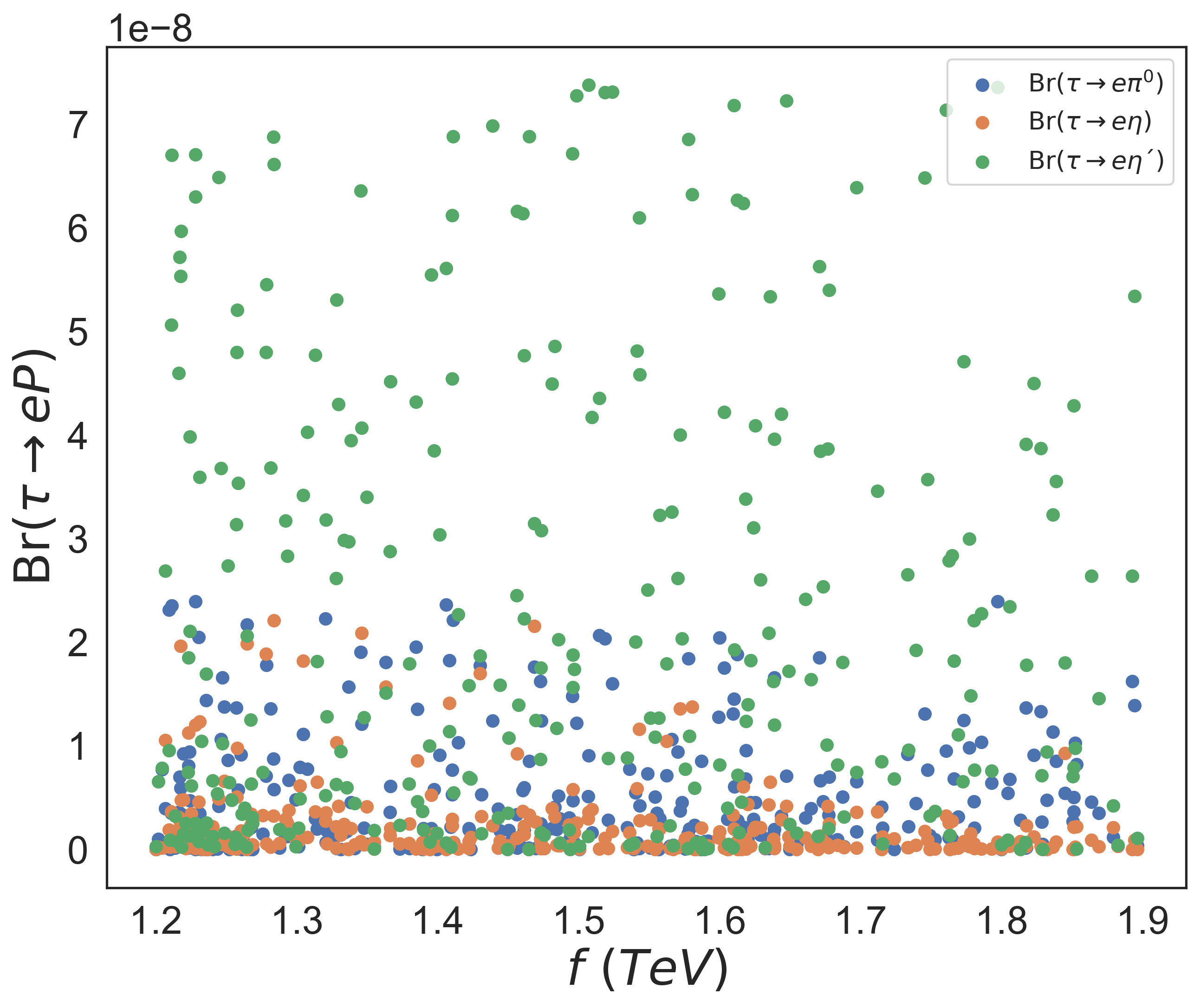}
        \caption{Scatter plot $f$ vs. $\mathrm{Br}(\tau \rightarrow e P)$ without Majorana neutrinos.}
        \label{fig: f_eP_sin_Majoranas}
    \end{minipage}
    \hspace{1cm}
    \begin{minipage}{0.45\linewidth}
    \centering
        \includegraphics[width=\linewidth]{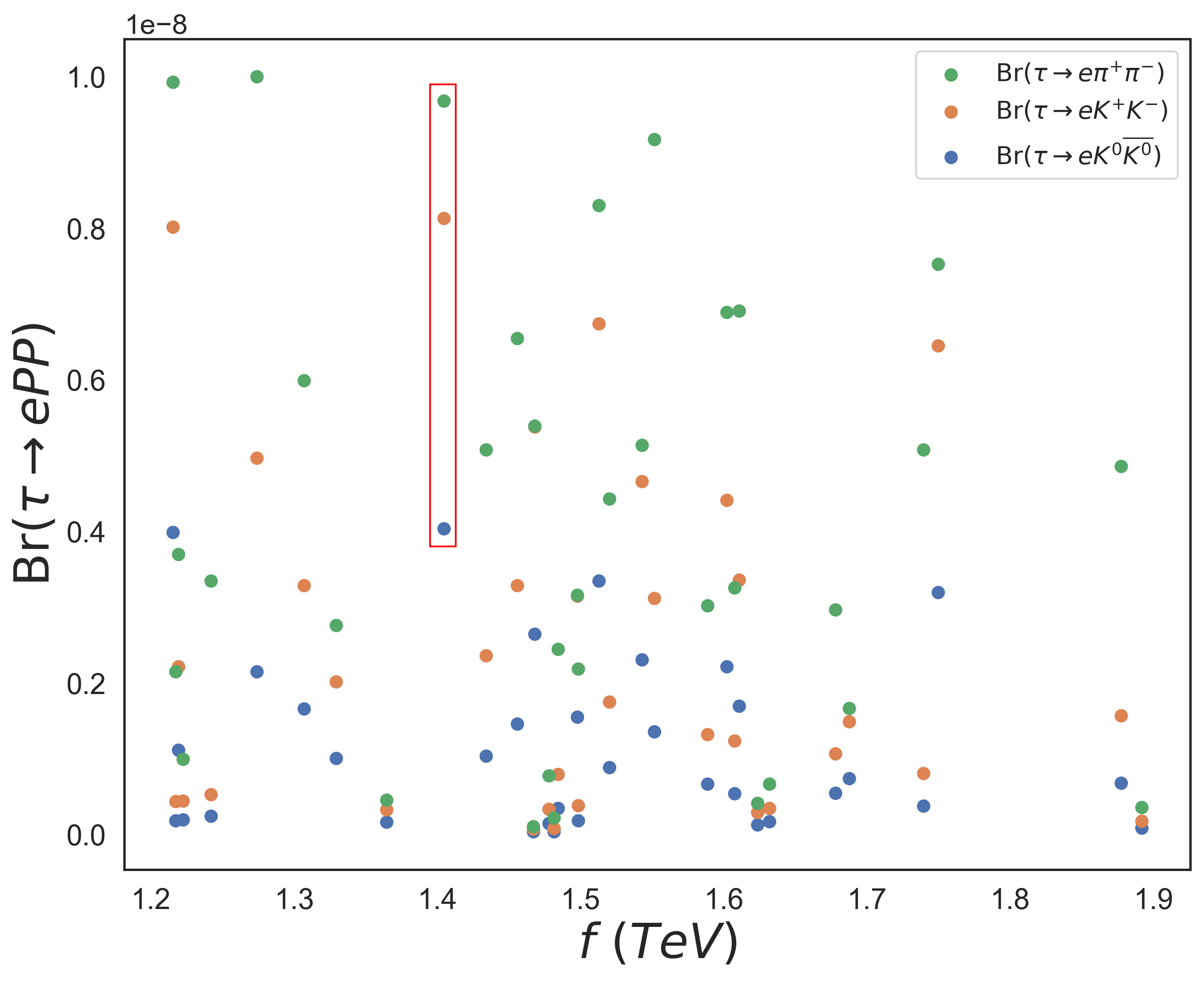}
        \caption{Scatter plot $f$ vs. $\mathrm{Br}(\tau \rightarrow e PP)$ without Majorana neutrinos contribution.}
        \label{fig: f_ebr_sin_Majorana}
    \end{minipage}
    \hspace{1cm}
    \begin{minipage}{0.45\linewidth}
    \centering
        \includegraphics[width=\linewidth]{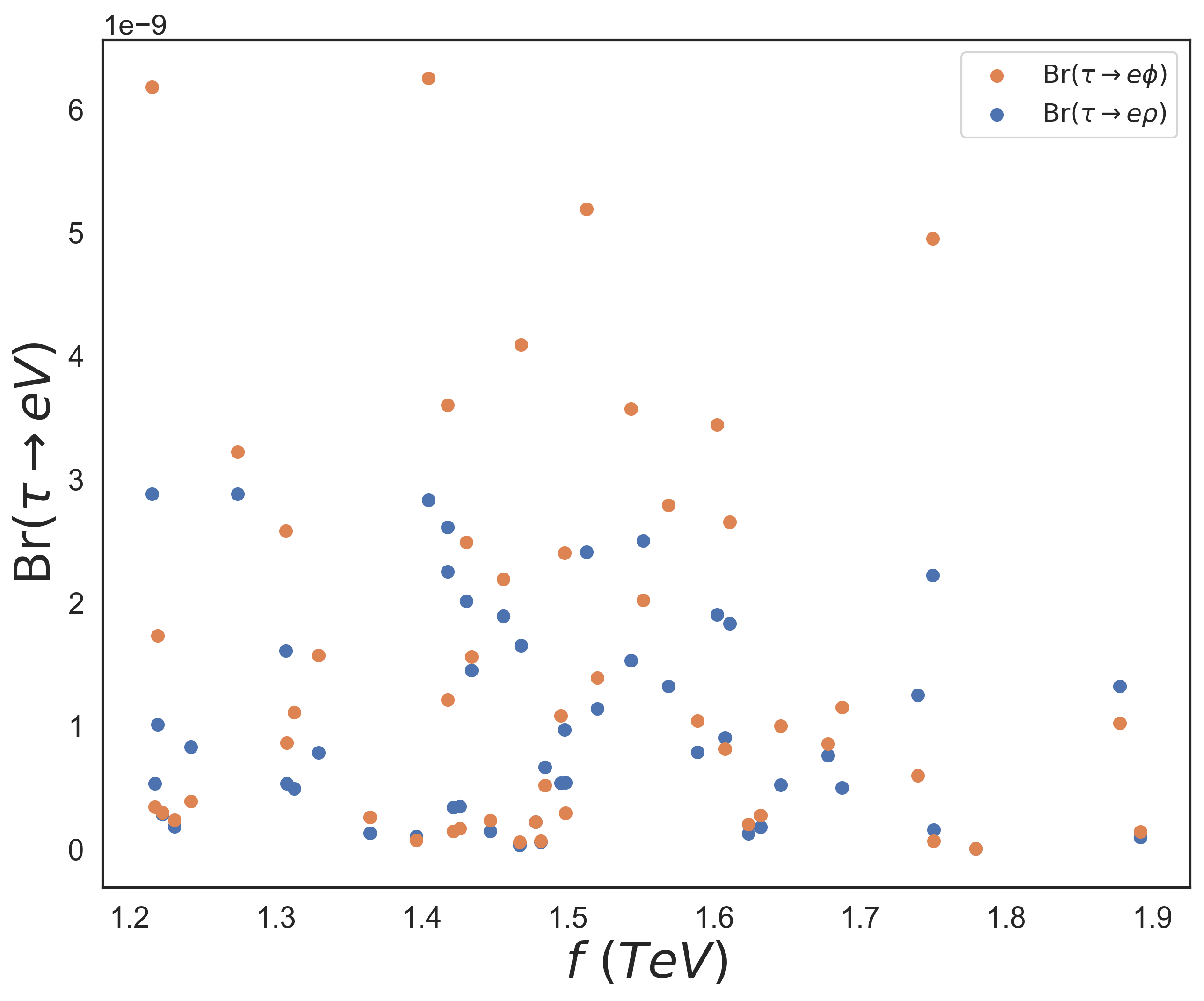}
        \caption{Scatter plot $f$ vs. $\mathrm{Br}(\tau \rightarrow e V)$ without Majorana neutrinos contribution.}
        \label{fig: f_ebrV_sin_Majorana}
    \end{minipage}
\end{figure}

In Figures \ref{fig: f_eP_sin_Majoranas}, \ref{fig: f_ebr_sin_Majorana} and \ref{fig: f_ebrV_sin_Majorana} we see how the branching ratios for the three decay modes with $\ell = e$ behave with respect to $f$ (similarly for $\ell = \mu$, not displayed). The decay mode with $ P = \eta'$ reaches the highest values, meanwhile the $P = \eta$ channel is the most restricted one among the $\tau \to \ell P$ processes.

We note that the branching ratios for $\tau \to e PP$ processes are arranged in triplets in Figure \ref{fig: f_ebr_sin_Majorana}, this pattern is just present in this type of processes (also for $\ell = \mu$), where the $\pi^+\pi^-$ mode (green points) has the largest probability, as  expected. The scatter plot in Figure \ref{fig: f_ebrV_sin_Majorana} shows us that the decay mode with $V = \phi$ as final state is larger than with $V = \rho$ (the same for $\ell = \mu$).

The predicted mean values of branching ratios from our numerical analyses are one ($\eta'$ modes) or at most two orders of magnitude smaller than the current limits \cite{PDG}, which is quite promising recalling that we considered only particles from LHT in this analysis.

\clearpage

\subsection{With Majorana Neutrinos Contribution}
\label{sec:With Majorana Neutrinos Contribution}

Now we add the contribution from Majorana neutrinos. 
Then, the decays can be distinguished by their neutral couplings: $(\theta S \theta^{\dagger})_{e \tau}-$processes and $(\theta S \theta^{\dagger})_{\mu \tau}-$processes. Both types of decays share almost the same free parameters just differing by the neutral couplings of heavy Majorana neutrinos. Therefore, the phenomenological analysis for $\tau \to \ell P \ (\ell = e,\mu)$ decays is done through a single Monte Carlo simulation in which the six decays are run simultaneously. Also, as in section \ref{sec: Without Majorana Neutrinos Contribution}, the $\tau \to \ell PP, \ \ell V$ processes are computed jointly in another Monte Carlo simulation. In Tables \ref{tau_ellP} and \ref{tau_ellPP} the corresponding analyses results are shown.

In this case, 
the NP scale is slightly larger ($\sim 25$ GeV) than without Majoranas, recalling that $f \sim 1.50$ TeV in section \ref{sec: Without Majorana Neutrinos Contribution}. 

T-odd particles, both leptons and quarks, are heavier now than in section \ref{sec: Without Majorana Neutrinos Contribution}, since in this case the masses mean values of T-odd leptons are always above 3 TeV and for T-odd quarks their masses  exceed 3 TeV, in contrast to results in Tables \ref{tau_ellP_sin_Majoranas} and \ref{tau_ellPP_sin_Majorana}. Actually, the mass ordering between T-odd leptons and quarks is reversed with respect to section \ref{sec: Without Majorana Neutrinos Contribution}.

\begin{table}[!ht]
    \scriptsize
    \centering
    \begin{tabular}{c|c|c|c}
    \hline \hline
    \multicolumn{4}{c}{\textbf{$\tau \to \ell P \ (\ell = e,\mu) \ (\mathrm{C.L.} = 90\%)$} with Majorana neutrinos contribution.} \\
    \hline \hline
    \multicolumn{2}{c|}{New physics (NP) scale (TeV)} & \multicolumn{2}{c}{Mixing angles} \\
    \hline
     $f$ & 1.51
     & $\theta_{V}$ & $43.07^{\circ}$ \\
     \cline{1-2}
    \multicolumn{2}{c|}{Branching ratio} & $\theta_{W}$ & $42.82^{\circ}$ \\
    \hline
     Br($\tau \to e \pi^{0}$) & $8.69
     \times 10^{-9}$  & \multicolumn{2}{|c}{Masses of partner leptons $(m_{\nu^{c}} = m_{\ell^{c}})$(TeV)} \\
     \cline{3-4}
     Br($\tau \to \mu \pi^{0}$) & $6.96
     \times 10^{-9}$ & $m_{\nu^{c}_{1}}$ & 3.26
     \\
     Br($\tau \to e \eta$) & $6.19
     \times 10^{-9}$ & $m_{\nu^{c}_{2}}$ & 3.26
     \\
     Br($\tau \to \mu \eta$) & $5.19
     \times 10^{-9}$ & $m_{\nu^{c}_{3}}$ & 3.30
     \\
     \cline{3-4}
     Br($\tau \to e \eta'$) & $2.19
     \times 10^{-8}$ & \multicolumn{2}{|c}{Masses of partner quarks $(m_{u^{c}} = m_{d^{c}})$ (TeV) } \\
     \cline{3-4}
     Br($\tau \to \mu \eta'$) & $1.94
     \times 10^{-8}$ & $m_{u^{c}_{i}}$ & 3.31
     \\
     \hline
     \multicolumn{2}{c|}{Masses of T-odd leptons (TeV)}& \multicolumn{2}{c}{Masses of heavy  Majorana  neutrinos (TeV)}\\
     \hline
     $m_{\ell^{1}_{H}}$ & 3.06
     & $\mathrm{M_{1}}$ & 19.18
     \\
     $m_{\ell^{2}_{H}}$ & 3.03
     & $\mathrm{M_{2}}$ & 19.07
     \\
     $m_{\ell^{3}_{H}}$ & 3.03
     & $\mathrm{M_{3}}$ & 19.25
     \\
     \cline{3-4}
     $m_{\nu^{1}_{H}}$ & 3.05
     & \multicolumn{2}{|c}{Neutral couplings of heavy Majorana neutrinos} \\
     \cline{3-4}
     $m_{\nu^{2}_{H}}$ & 3.02
     & $|(\theta S \theta^{\dagger})_{e \tau}|$ & $3.32
     \times 10^{-7}$\\
     $m_{\nu^{3}_{H}}$ & 3.02
     & $|(\theta S \theta^{\dagger})_{\mu \tau}|$ & $3.90
     \times 10^{-7}$ \\
     \hline
     \multicolumn{2}{c|}{Masses of T-odd quarks (TeV)}&  & \\
     \cline{1-2}
     $m_{d^{i}_{H}}$ & 2.78
     &  & \\
     $m_{u^{i}_{H}}$ & 2.78
     &  & \\
     \hline
    \end{tabular}
    \caption{Mean values for branching ratios, masses of
    LHT heavy particles, mixing angles and neutral couplings  obtained by Monte Carlo simulation of $\tau \to \ell P \ (\ell = e,\mu)$ processes (case with Majorana neutrinos).}
    \label{tau_ellP}
\end{table}

\begin{table}[!ht]
    \scriptsize
    \centering
    \begin{tabular}{c|c|c|c}
    \hline \hline
    \multicolumn{4}{c}{\textbf{$\tau \to \ell PP, \ \ell V \ (\ell = e,\mu) \ (\mathrm{C.L.} = 90\%)$} with Majorana neutrinos contribution} \\
    \hline \hline
    \multicolumn{2}{c|}{New physics (NP) scale (TeV)} & \multicolumn{2}{c}{Mixing angles} \\
    \hline
     $f$ & 1.54
     & $\theta_{V}$ & $43.61^{\circ}$ \\
     \cline{1-2}
    \multicolumn{2}{c|}{Branching ratio} & $\theta_{W}$ & $42.21^{\circ}$ \\
    \hline
     Br($\tau \to e \pi^{+}\pi^{-}$) & $4.75
     \times 10^{-9}$  & \multicolumn{2}{|c}{Masses of partner leptons $(m_{\nu^{c}} = m_{\ell^{c}})$(TeV)} \\
     \cline{3-4}
     Br($\tau \to \mu \pi^{+}\pi^{-}$) & $4.90
     \times 10^{-9}$ & $m_{\nu^{c}_{1}}$ & 2.91
     \\
     Br($\tau \to e K^{+}K^{-}$) & $2.53
     \times 10^{-9}$ & $m_{\nu^{c}_{2}}$ & 2.99
     \\
     Br($\tau \to \mu K^{+}K^{-}$) & $3.38
     \times 10^{-9}$ & $m_{\nu^{c}_{3}}$ & 2.95
     \\
     \cline{3-4}
     Br($\tau \to e K^{0} \overline{K^{0}}$) & $1.16
     \times 10^{-9}$ & \multicolumn{2}{|c}{Masses of partner quarks $(m_{u^{c}} = m_{d^{c}})$ (TeV) } \\
     \cline{3-4}
     Br($\tau \to \mu K^{0} \overline{K^{0}}$) & $1.50
     \times 10^{-9}$ & $m_{u^{c}_{i}}$ & 3.08
     \\
     \cline{3-4}
     Br($\tau \to e \rho$) & $1.33
     \times 10^{-9}$ & \multicolumn{2}{c}{Masses of heavy  Majorana  neutrinos (TeV)} \\
     \cline{3-4}
     Br($\tau \to \mu \rho$) & $1.39
     \times 10^{-9}$ & $\mathrm{M_{1}}$ & 18.82
     \\
     Br($\tau \to e \phi$) & $1.78
     \times 10^{-9}$ & $\mathrm{M_{2}}$ & 19.33
     \\
     Br($\tau \to \mu \phi$) & $2.08
     \times 10^{-9}$ & $\mathrm{M_{3}}$ & 18.92
     \\
     \cline{1-2}
     \hline
     \multicolumn{2}{c|}{Masses of T-odd leptons (TeV)}& \multicolumn{2}{|c}{Neutral couplings of heavy Majorana neutrinos}\\
     \hline
     $m_{\ell^{1}_{H}}$ & 3.49
     & $|(\theta S \theta^{\dagger})_{e \tau}|$ & $2.20
     \times 10^{-7}$
     \\
     $m_{\ell^{2}_{H}}$ & 3.47
     & $|(\theta S \theta^{\dagger})_{\mu \tau}|$ & $3.14
     \times 10^{-7}$
     \\
     \cline{3-4}
     $m_{\ell^{3}_{H}}$ & 3.21
     &  & 
     \\
     $m_{\nu^{1}_{H}}$ & 3.48
     & &  \\
     $m_{\nu^{2}_{H}}$ & 3.46
     &  & \\
     $m_{\nu^{3}_{H}}$ & 3.20
     &  &  \\
     \cline{1-2}
     \multicolumn{2}{c|}{Masses of T-odd quarks (TeV)}&  & \\
     \cline{1-2}
     $m_{d^{i}_{H}}$ & 3.73
     &  & \\
     $m_{u^{i}_{H}}$ & 3.71
     &  & \\
     \hline
    \end{tabular}
    \caption{Mean values for branching ratios, masses of
    LHT heavy particles, mixing angles and neutral couplings obtained by Monte Carlo simulation of $\tau \to \ell PP, \ \ell V \ (\ell = e,\mu)$ processes.}
    \label{tau_ellPP}
\end{table}

Only for $\tau \to \ell PP, \ \ell V$ processes with Majorana neutrinos, the partner leptons have masses below 3 TeV. Regarding the mean values for the lepton quark masses, they are always heavier without  Majorana neutrinos. In this case their maximum value is $ \sim 3.31$ TeV in Table \ref{tau_ellP}. Meanwhile, in Table \ref{tau_ellP_sin_Majoranas}, they have mean masses $\sim 3.55$ TeV.

We see that the presence of Majorana neutrinos does not change sizeably the mean values for the mixing angles $\theta_{V}$ and $\theta_{W}$, thus they can be considered practically equal to those from Subsection \ref{sec: Without Majorana Neutrinos Contribution}, nearly maximal.



\begin{figure}[!ht]
     \centering
     \includegraphics[scale=0.48]{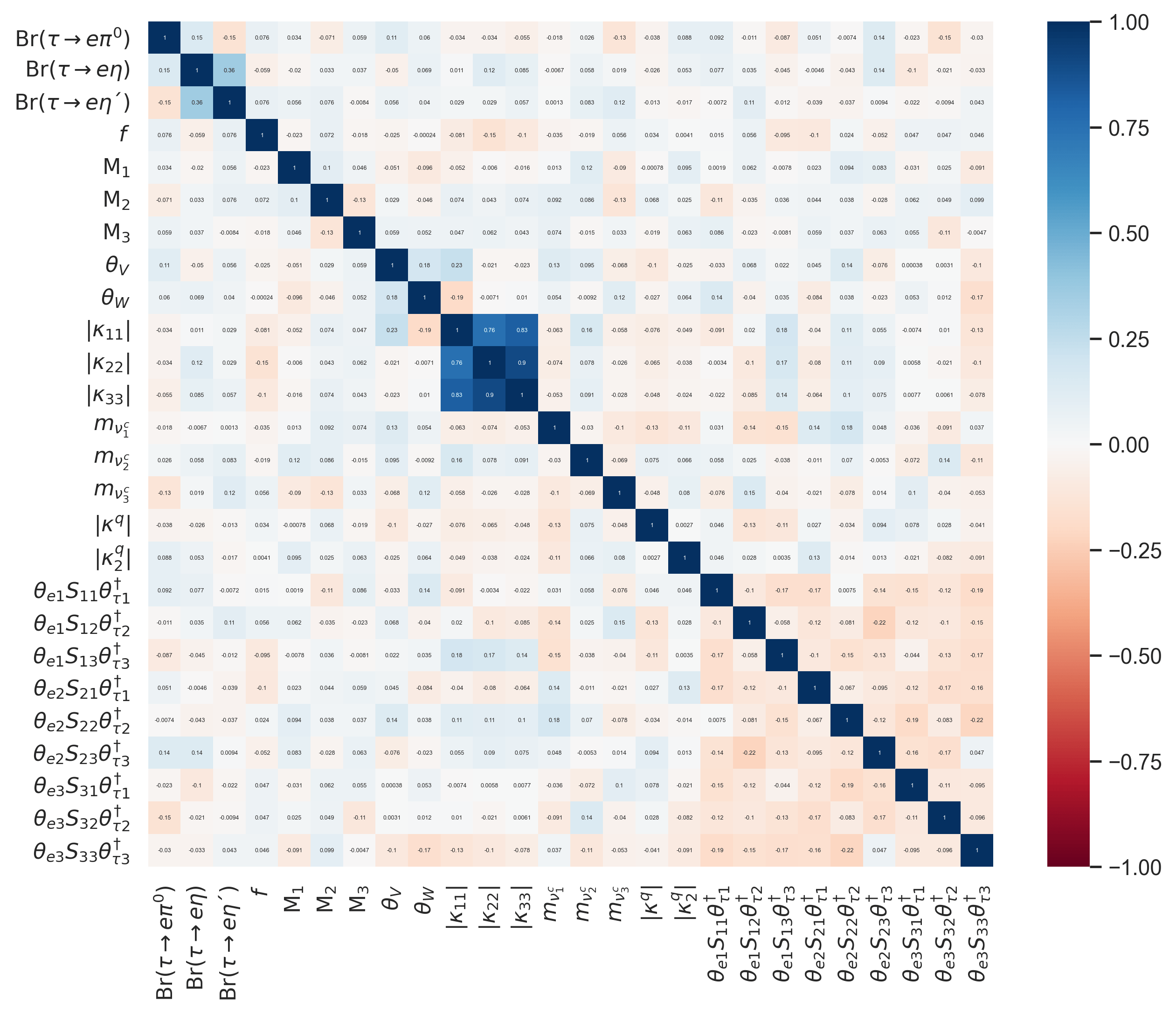}
     \caption{Heat map showing absence of sizable correlations  among $\tau \to e P$ decays and their free parameters (case with Majorana neutrinos).}
     \label{fig:tau_eP}
 \end{figure}
 
 \begin{figure}[!ht]
     \centering
     \includegraphics[scale=0.48]{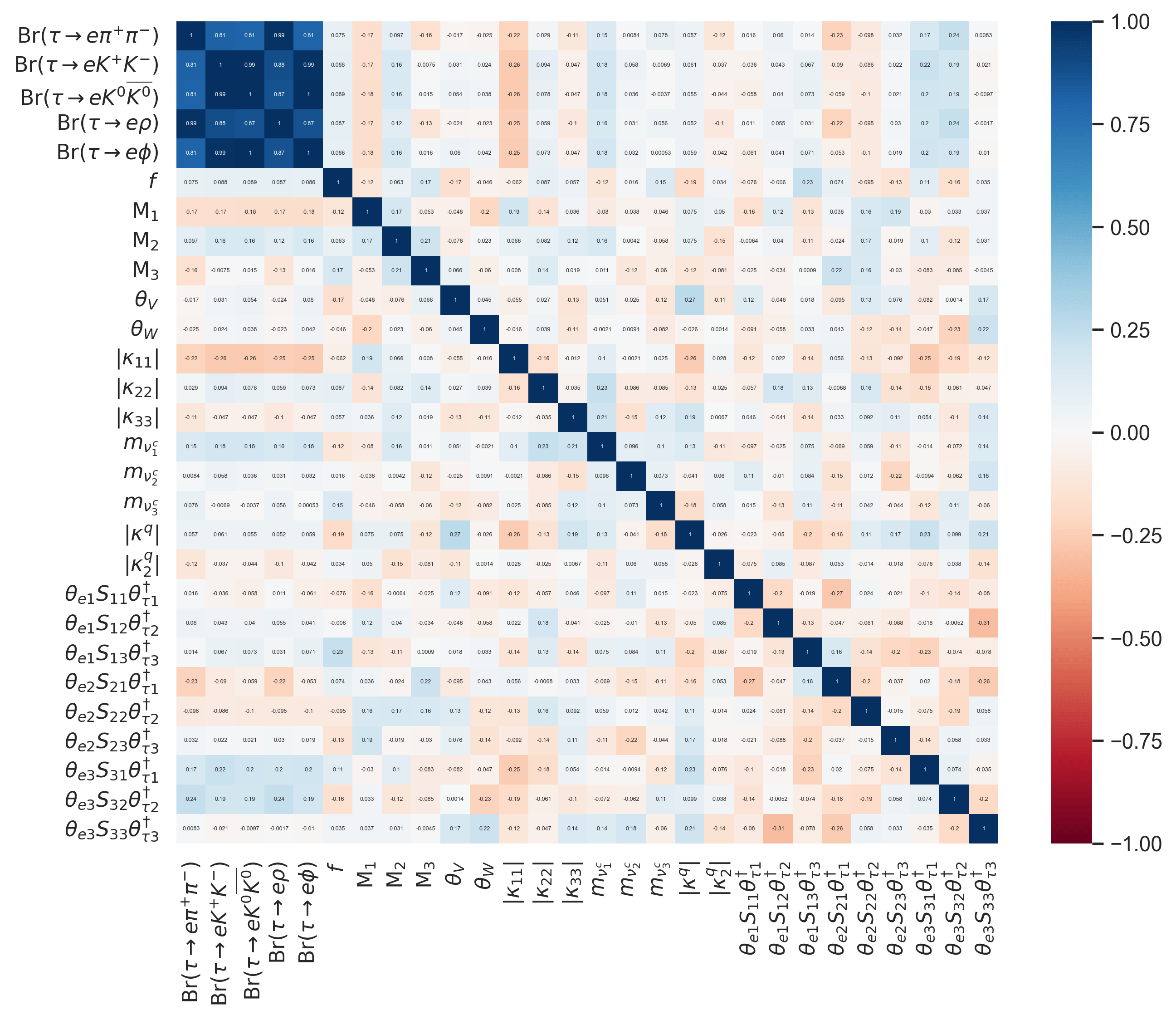}
     \caption{Heat map for $\tau \to ePP, \ eV$ decays and their free parameters considering Majorana neutrinos.}
     \label{fig:tau_ePP_V}
\end{figure}

The novel parameters that appear in these analyses are the masses of heavy Majorana neutrinos and their neutral couplings. We observe that these masses  are $\sim 19$ TeV (mean value for each heavy Majorana neutrino)
. Their neutral couplings $|(\theta S \theta^{\dagger})_{\ell \tau}| \ (\ell = e, \mu)$ for both analyses ($\tau \to \ell P$ and $\tau \to \ell PP, \ \ell V$ processes) have the same order of magnitude,  $\sim\mathcal{O}(10^{-7})$, in agreement with our previous paper \cite{Pacheco}.

In the following heat maps, Figures \ref{fig:tau_eP} and \ref{fig:tau_ePP_V}, branching ratios with $\ell = e$ are almost uncorrelated with their free parameters (similarly for $\ell = \mu$). 

The correlations among Yukawa couplings of T-odd leptons is kept from the previous analysis,  section \ref{sec: Without Majorana Neutrinos Contribution}
. 

 
The correlations among branching fractions looks different than in section \ref{sec: Without Majorana Neutrinos Contribution}. For both processes, with $\ell = e,\mu$, including Majorana neutrinos contribution, the decay modes with $P = \eta$ and $P = \eta'$ have the highest correlations. To explain that, we need to realize that their $a_{P}^{k}$ and $b_{P}^{k}$ factors (see appendix C) are numerically similar in these cases.

Correlation among branching ratios for $\tau \to \ell PP, \ \ell V$ processes look quite alike in all cases. This can be understood as a result of the largest contribution coming always from the pions loop in the  $F_{V}^{PP}(s)$ function. This causes the correlations among them to be maximal.


In contrast to our previous analysis \cite{Pacheco}, here the heavy Majorana neutrinos are barely correlated among them. In Ref.~\cite{Pacheco}, the mean value for heavy Majorana masses is around $17.2$ TeV, differing slightly $(\sim 0.12\%)$ in all cases, which is lighter (by $\sim 2$ TeV) than in this work. 

\begin{figure}[!ht]
\centering
    \begin{minipage}{0.45\linewidth}
    \centering
        \includegraphics[width=\linewidth]{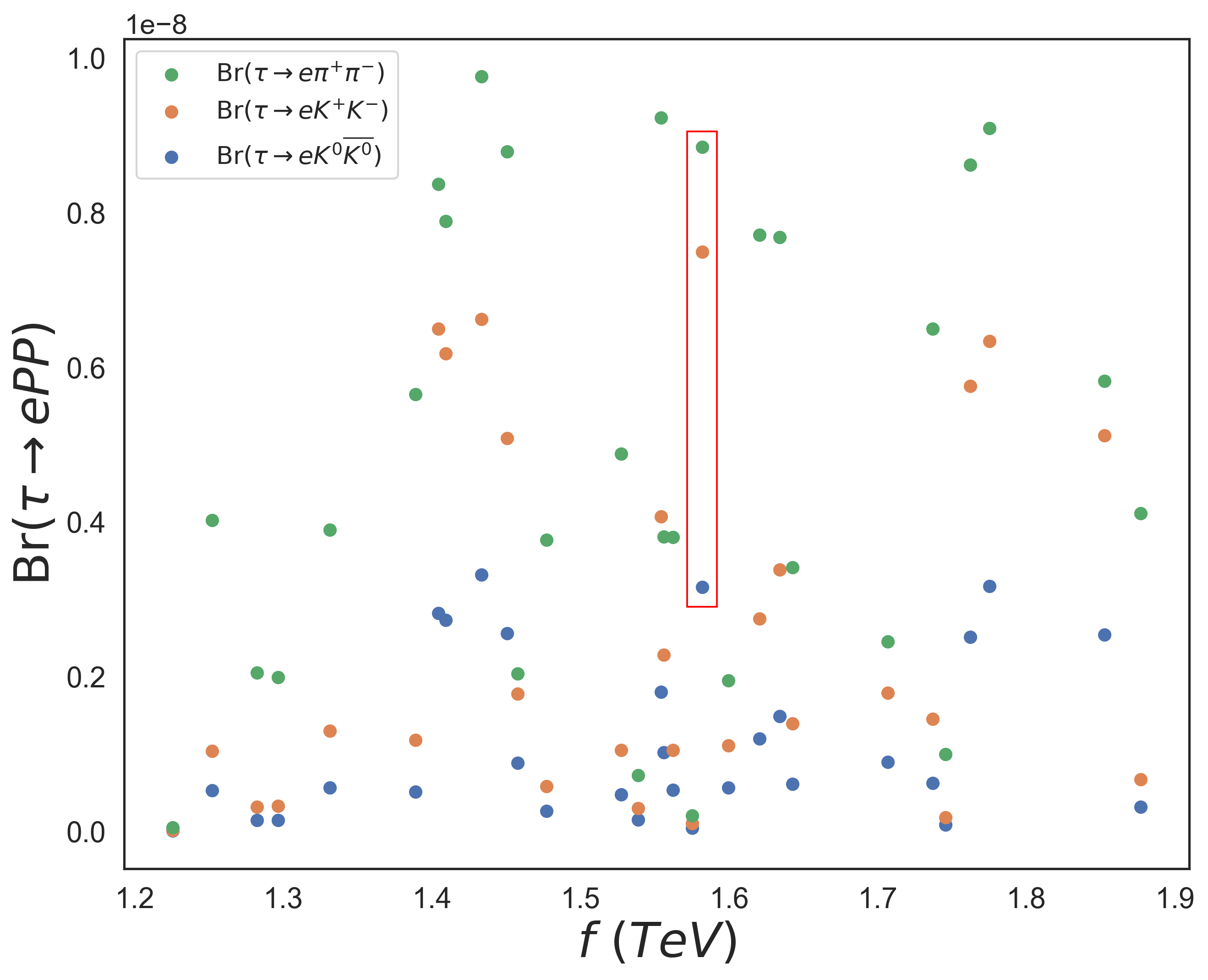}
        \caption{Scatter plot $f$ vs. $\mathrm{Br}(\tau \rightarrow e PP)$ considering Majorana neutrinos.}
        \label{fig: f_ebr}
    \end{minipage}
    \hspace{1cm}
    \begin{minipage}{0.45\linewidth}
        \centering
        \includegraphics[width=\linewidth]{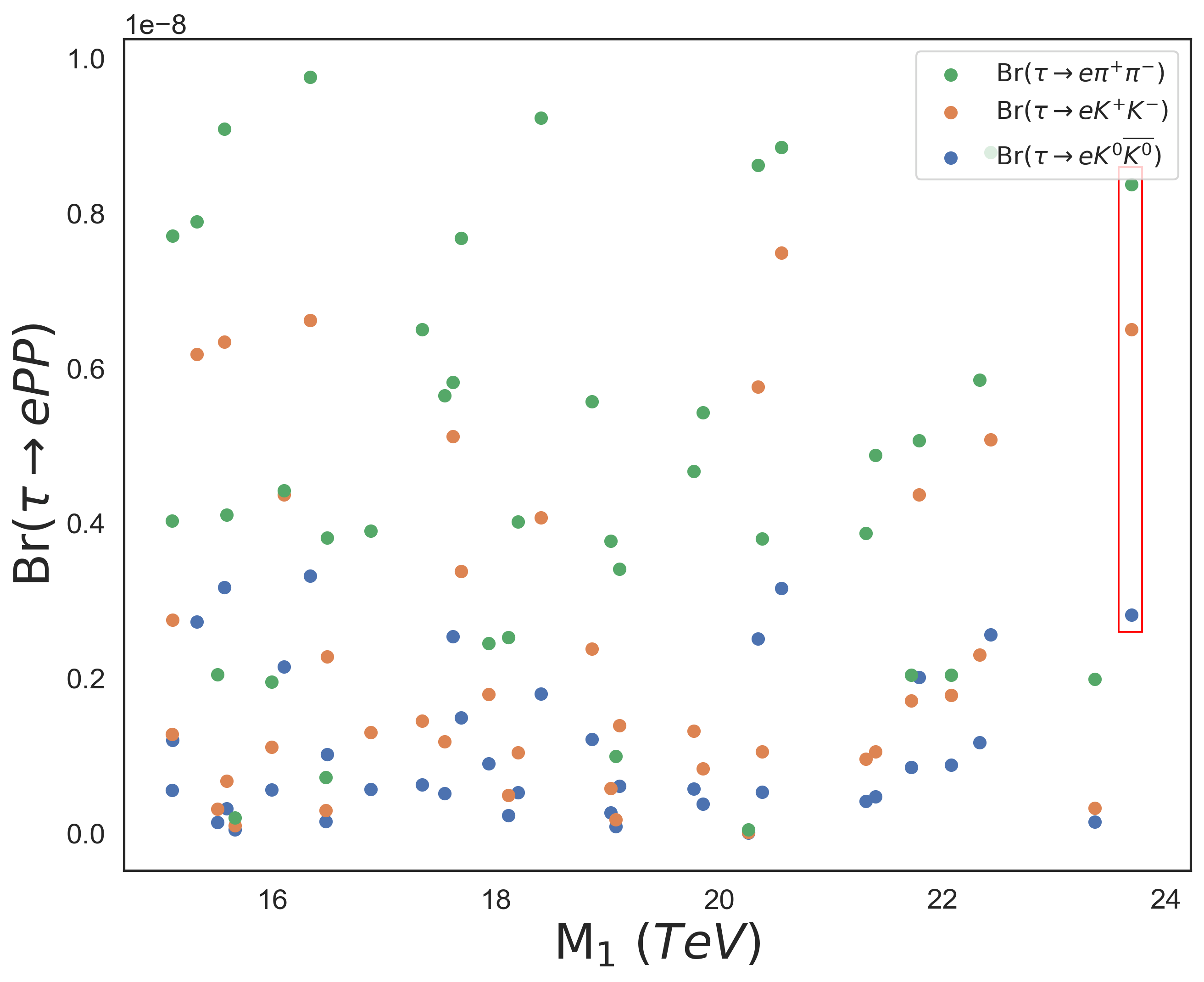}
        \caption{Scatter plot $\mathrm{M}_{1}$ vs. $\mathrm{Br}(\tau \rightarrow e PP)$ considering Majorana neutrinos.}
        \label{fig: M1_ebr}
    \end{minipage}
\end{figure}

We include only a few scatter plots, Figures \ref{fig: f_ebr} and \ref{fig: M1_ebr}, corresponding to $\tau \to e PP$ (similarly for $\ell = \mu$) processes since, as in section \ref{sec: Without Majorana Neutrinos Contribution}, the pattern of triads appears again. The $\pi^{+}\pi^{-}$ is the most probable decay channel, as  expected. 

In this case for $\tau \to \ell P$ and $\tau \to \ell V$ processes, the interpretation of scatter plots is very similar to the ones in section \ref{sec: Without Majorana Neutrinos Contribution}, so they are not  shown. 

Now, additionally, we include the scatter plot between branching ratio versus mass of Majorana neutrinos in Figure \ref{fig: M1_ebr}, that looks similar to the branching ratio versus $f$ plot. This was expected because both magnitudes are related, $M_i\leq 4 \pi f$.  For the reasons commented above, we do not include the scatter plots corresponding to $\tau \to \ell P$ and $\tau \to \ell V$ processes, but it is straightforward to visualize their behaviors (similarly for $\ell = \mu$).

Our branching ratios are slightly larger when the Majorana neutrinos contribution is included. Again, results are more promising for the $\tau\to\ell\eta'$ modes, which are only one order of magnitude smaller (at most two, in some other decay channels) than current upper limits \cite{PDG}.

\begin{figure}
    \centering
    \includegraphics[scale = 0.5]{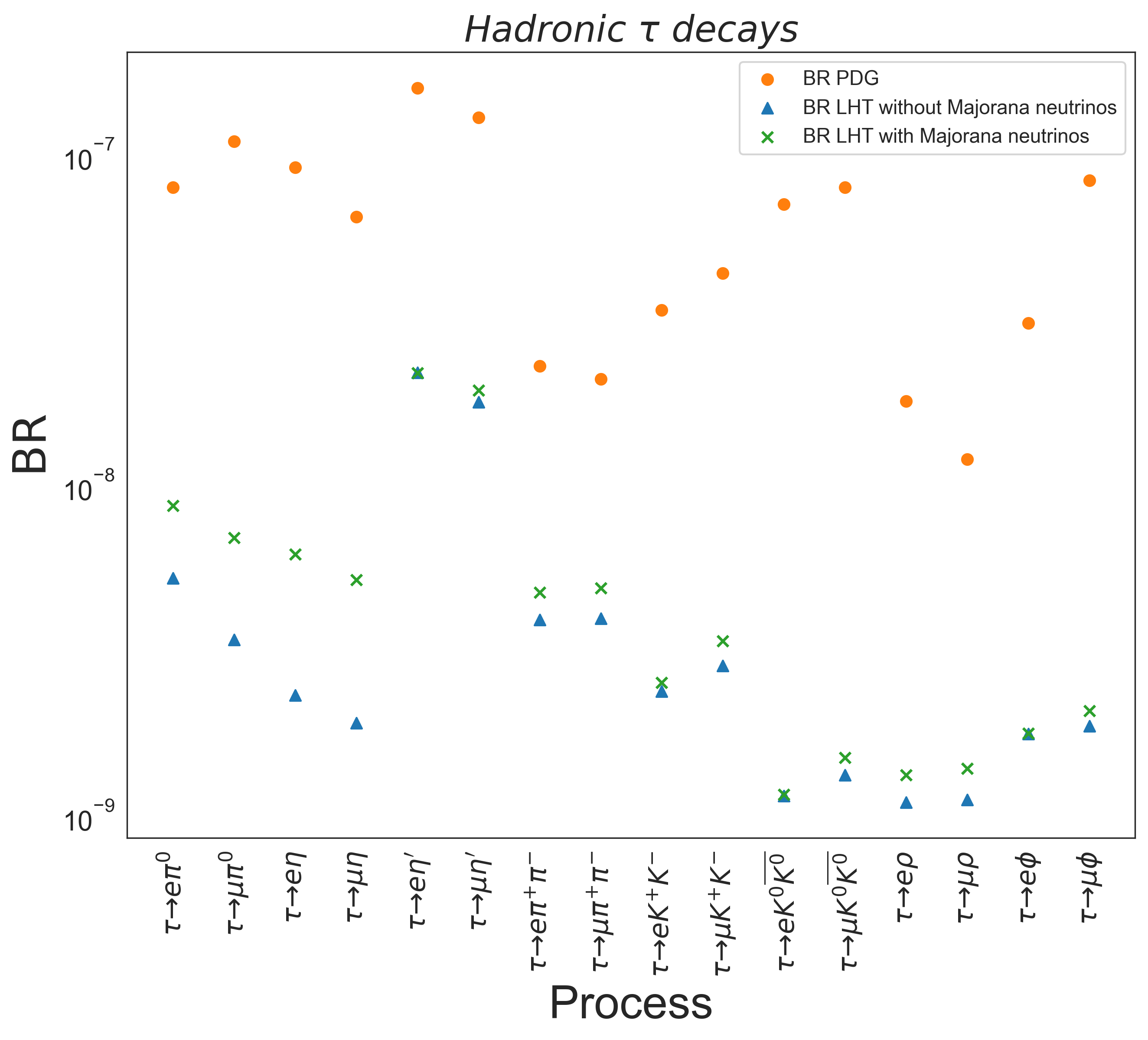}
    \caption{Scatter plot comparing our predicted mean values with the upper limits from PDG \cite{PDG}.}
    \label{fig:results_vs_PDGl}
\end{figure}

Altogether, the scatter plot displayed in Figure \ref{fig:results_vs_PDGl} compares  the predicted mean values of branching ratios in our model with (green marks) and without (blue marks) Majorana contributions  with the upper limits from the PDG \cite{PDG} (orange marks). 

\clearpage
 
\section{Conclusions}\label{sec:Concl}

Although LFV processes have been studied extensively within the LHT model, an analysis of semileptonic tau decays was still lacking. We have tackled it here, both considering only effects of $T$-odd and partner fermions, and also adding the contributions from Majorana neutrinos realizing an inverse seesaw mechanism of type I. Our main results, according to the mean values of our simulations, are summarized in the following:

\begin{itemize}
    \item Masses of particles coming from LHT, T-odd and partner fermions, are below $4$ TeV,  almost $5$ times lighter than heavy Majorana neutrinos. This is related with the next item.
    \item The new physics (NP) scale is around $f \sim 1.5$ TeV, which is accompanied by heavy Majorana neutrinos with $\mathrm{M}_{i} \sim 19$ TeV (we recall that $\mathrm{M}_{i} \sim4\pi f$). Compared to our previous work \cite{Pacheco},  these masses of heavy Majorana neutrinos  are heavier by $\sim 2$ TeV ($\sim 10 \%$ of difference) and $f$ is fully consistent.
    \item The magnitudes of neutral couplings of heavy Majorana neutrinos $|(\theta S \theta^{\dagger})_{\ell \tau}| \ (\ell = e,\mu)$ match those  reported in  ref.~\cite{Pacheco}.
    \item The masses of T-odd particles from $\tau \to \ell P$ processes differ from those in the $\tau \to \ell PP ,\, \ell V \ (\ell = e,\mu)$ decays by $\sim 1$ TeV. All other mean values from Tables \ref{tau_ellP_sin_Majoranas}, \ref{tau_ellPP_sin_Majorana}, \ref{tau_ellP}, and \ref{tau_ellPP} agree. Taking this into account, we consider these values as representative for the present analysis of our model. A global analyses (including purely leptonic processes and conversions in nuclei, see e.g. refs.~\cite{delAguila:2010nv,Husek:2020fru,Ramirez:2022zpk}) is required, however, and will be presented elsewhere.
    \item All our results are very promising and will be probed in future measurements, as they lie approximately only one order of magnitude below currents bounds \cite{PDG}.
\end{itemize}

\section*{Acknowledgements}
We are indebted to Jorge Portol\'es for useful discussions. We are also extremely thankful  to the anonymous referee for thoroughly checking the paper and giving very  useful suggestions for improving its presentation. I. P. acknowledges Conacyt funding his Ph. D and P. R. the financial
support of C\'atedras Marcos Moshinsky (Fundaci\'on Marcos Moshinsky) and Conacyt's project within ‘Paradigmas y Controversias de la Ciencia 2022’, number 319395.

\appendix
\section{Appendix: Form Factor Functions of T-odd Contribution}
\label{App: A}

The functions involved in the $F_{M}^{\gamma}$ form factor of T-odd contributions are
 (we omit their second argument, which is always $Q^2$, below)
\begin{equation}
    \begin{split}
        F_{Z_{H}}(x) & = -\frac{1}{3} + \frac{2x + 5x^{2}-x^{3}}{8(1-x)^{3}} + \frac{3x^{2}}{4(1-x)^{4}}\ln x \\
        & + \frac{1}{48} \frac{Q^{2}}{M_{Z_{H}}^{2}} \left( \frac{24 + 546x - 2479x^{2} + 561x^{3} - 339x^{4} + 67x^{5}}{60(1-x)^{5}} + \frac{x(6-17x-16x^{2})\ln x}{(1-x)^{6}} \right), \\
        F_{W_{H}}(x) & = \frac{5}{6} - \frac{3x - 15x^{2}-6x^{3}}{12(1-x)^{3}} + \frac{3x^{3}}{2(1-x)^{4}}\ln x \\
        & + \frac{x}{24} \frac{Q^{2}}{M_{W_{H}}^{2}} \left( \frac{134 - 759x + 1941x^{2} - 2879x^{3} - 69x^{4} + 12x^{5} }{60(1-x)^{5}} + \frac{x^{3} (4 - 34x + 3x^{3}) \ln x}{(1-x)^{6}} \right), \\ 
        F_{\tilde{\nu}^{c}}(x) & = \frac{-1 + 5x + 2x^{2}}{12(1-x)^{3}} + \frac{x^{2}}{2(1-x)^{4}}\ln x \\
        & + \frac{x^{2}}{24} \frac{Q^{2}}{M_{\Phi}^{2}} \left( \frac{13 - 87x + 333x^{2} + 293x^{3} - 12x^{4}}{60(1-x)^{5}} + \frac{x^{3}(8+x)\ln x}{(1-x)^{6}} \right), \\
        F_{\tilde{\ell}^{c}}(x) & = \frac{-4 + 5x + 5x^{2}}{6(1-x)^{3}} - \frac{x(1-2x)}{(1-x)^{4}} \ln x \\
        & + \frac{1}{8} \frac{Q^{2}}{M_{\Phi}^{2}} \left( \frac{4 + 129x - 231x^{2} - 91x^{3} + 9x^{4}}{60(1-x)^{5}} + \frac{x(1-4x^{2})\ln x}{(1-x)^{6}} \right),
    \end{split}
    \label{eq7.1.9}
\end{equation}
with $a = \frac{M_{W_{H}}^{2}}{M_{A_{H}}^{2}} = \frac{5c_{W}^{2}}{s^{2}_{W}}\sim15$ and $M_{W_{H}} = M_{Z_{H}}$.
We have used $M_{W}^{2}/M_{W_{H}}^{2} = \upsilon^{2}/(4f^{2})$.

In turn, the functions from eq. (\ref{eq7.1.11}) read
\begin{equation}
    \begin{split}
        G_{Z}^{(1)}(x) & = \frac{1}{36}+ \frac{x(18-11x-x^{2})}{48(1-x)^{3}}-\frac{4-16x+9x^{2}}{24(1-x)^{4}}\ln x,\\
        G_{W}^{(1)}(x) & =-\frac{5}{18}+ \frac{x(12+x-7x^{2})}{24(1-x)^{3}}+\frac{x^{2}(12-10x+x^{2})}{12(1-x)^{4}}\ln x,\\
        G_{\tilde{\nu}^{c}}^{(1)}(x) & = \frac{2-7x+11x^{2}}{72(1-x)^{3}}+\frac{x^{3}}{12(1-x)^{4}}\ln x, \\
        G_{\tilde{\ell}^{c}}^{(1)}(x) & = \frac{20-43x+29x^{2}}{36(1-x)^{3}}+\frac{2-3x+2x^{3}}{6(1-x)^{4}}\ln x.
    \end{split}
    \label{eq7.1.12}
\end{equation}

Form factor functions of $F_{L}^{Z}$ are
\begin{align}
        H_{L}^{W(0)}(x) & = \frac{6-x}{1-x} + \frac{2+3x}{(1-x)^{2}}\ln x, \nonumber \\
        H_{L}^{W}(x) & = 2G_{Z}^{(1)}(x) - 2c_{W}^{2}G_{W}^{(1)}(x), \nonumber \\
        H_{L}^{A/Z}(x) & = G_{Z}^{(1)}(x), \nonumber \\
        H_{L}^{\tilde{\nu}}(x) & = \frac{1}{2}G_{\tilde{\ell}^{c}}^{(1)}(x) - 2c_{W}^{2}G_{\tilde{\nu}^{c}}^{(1)}(x), \nonumber \\
        H_{L}^{\tilde{\ell}}(x) & = G_{\tilde{\ell}^{c}}^{(1)}(x).
    \label{eq7.1.17}
\end{align}
The expression for $G_{Z}^{(1)}$, $G_{W}^{(1)}$, $G_{\tilde{\nu}^{c}}^{(1)}$ and $G_{\tilde{\ell}^{c}}^{(1)}$ can be consulted from eq. (\ref{eq7.1.12}).

The four-point functions that appear in the box form factor, in the terms of the mass ratios $x = m_{1}^{2}/m_{0}^{2}$, $y = m_{2}^{2}/m_{0}^{2}$, $z = m_{3}^{2}/m_{0}^{2}$, turn out to be
\begin{equation}
    \begin{split}
        d_{0}(x,y,z) \equiv m_{0}^{4}D_{0} = & \left[ \frac{x\ln x}{(1-x)(x-y)(x-z)} - \frac{y\ln y}{(1-y)(x-y)(y-z)} \right. \\
        & \left. + \frac{z\ln z}{(1-z)(x-z)(y-z)} \right],
    \end{split} 
    \label{C04}
\end{equation}
\begin{equation}
    \begin{split}
        \Bar{d}_{0}(x,y,z) \equiv 4m_{0}^{2}D_{00} = & \left[ \frac{x^{2}\ln x}{(1-x)(x-y)(x-z)} - \frac{y^{2}\ln y}{(1-y)(x-y)(y-z)} \right. \\
        & \left. + \frac{z^{2}\ln z}{(1-z)(x-z)(y-z)} \right],
    \end{split} 
    \label{C05}
\end{equation}
\begin{equation}
        \Bar{d}'_{0}(x,y,z) =  \frac{x^{2}\ln x}{(1-x)(x-y)(z-x)} + \frac{y^{2}\ln y}{(1-y)(x-y)(z-y)} + \frac{z^{2}\ln z}{(1-z)(x-z)(y-z)},
    \label{C06}
\end{equation}
with $\Bar{d}(x,y) = \Bar{d}'(x,y,1)$. For two equals masses $(m_{0}=m_{3})$ we get
\begin{equation}
        d_{0}(x,y) = -\left[ \frac{x\ln x}{(1-x)^{2}(x-y)} - \frac{y\ln y}{(1-y)^{2}(x-y)} + \frac{1}{(1-x)(1-y)} \right],
    \label{C07}
\end{equation}
\begin{equation}
        \Bar{d}_{0}(x,y) =- \left[ \frac{x^{2}\ln x}{(1-x)^{2}(x-y)} - \frac{y^{2}\ln y}{(1-y)^{2}(x-y)} + \frac{1}{(1-x)(1-y)} \right].
    \label{C08}
\end{equation}

\section{Appendix: Form Factor Functions of Majorana Contribution}
\label{App: B}

The functions that make up the eq. (\ref{eq7.1.43}) are

\begin{align}
        F_{M}^{\chi^{h}}(x, Q^{2}) & = \frac{1}{3} - \frac{2x^{3}-7x^{2}+11x}{4(1-x)^{3}} + \frac{3x}{2(1-x)^{4}} \ln x, \nonumber \\
        & - \frac{1}{24}\frac{Q^{2}}{M_{j}^{2}} \left( \frac{134x_{j}^{5} - 759x_{j}^{4} + 1941x_{j}^{3} - 2879x_{j}^{2} - 69x_{j} + 12}{60x_{j}(1-x_{j})^{5}} + \frac{(4x_{j}^{2} - 34x_{j} + 3)\ln x_{j}}{(1-x_{j})^{6}} \right), \nonumber \\ 
        F_{L}^{\chi^{h}}(x,Q^{2}) & = 2\Delta_{\epsilon} + \frac{Q^{2}}{M_{j}^{2}} \left( - \frac{(12x^{2}-10x+1) \ln x}{6x(1-x)^{4}} + \frac{20x^{3}-96x^{2}+57x+1}{36x(1-x)^{3}}\right),
        \label{eq7.1.44}
\end{align}
with $\mathrm{M}_{j}$ the heavy Majorana neutrino masses and $\Delta_{\epsilon} = \frac{1}{\epsilon} - \gamma_{E} + \mathrm{ln}(4\pi) + \mathrm{ln}\left( \frac{\mu^{2}}{M_{W}^{2}} \right)$ which regulates the ultraviolet divergence in $4-2\epsilon$ dimensions, that is canceled by unitarity of mixing matrices (GIM-like mechanism). The explicit functions $F^{h}$, $G^{h}$, and $H^{h}$ are written as follows
{\small
\begin{align}
    F^{h}(x_{i};Q^{2}) & \approx - \frac{5}{2} \ \mathrm{ln} x_{i} + \frac{Q^{2}}{M_{j}^{2}} \left( \frac{1}{12x_{i}} (1-2s_{W}^{2}) \ \mathrm{ln} x_{i} \right), \nonumber \\
    G^{h}(x_{i},x_{j};M_{Z}^{2}) & = \frac{1}{2} \left( \Delta_{\epsilon}-\frac{1}{2} \right) - \frac{1}{2(x_{i}-x_{j})} \left(- \frac{(1-x_{j})\ \mathrm{ln}x_{i}}{(1-x_{i})} + \frac{(1-x_{i}) \ \mathrm{ln}x_{j}}{(1-x_{j})} \right)+ \frac{M_{Z}^{2}}{M_{j}^{2}} \times (\mathrm{terms}), \nonumber \\
    H^{h}(x_{i},x_{j};M_{Z}^{2}) & = - \frac{1}{4}\left( \Delta_{\epsilon} + \frac{1}{2} \right) - \frac{1}{4(x_{i} - x_{j})} \left( -\frac{(1-4x_{i})x_{j} \ \mathrm{ln}x_{i}}{(1-x_{i})} + \frac{(1 - 4x_{j})x_{i} \ \mathrm{ln}x_{j}}{(1-x_{j})} \right)+\frac{M_{Z}^{2}}{M_{i}^{2}} \times (\mathrm{terms}).
    \label{eq4.2.12}
\end{align}
}
where
{\small
\begin{align}
    G^{h}(x_{i},x_{j};Q^{2}): \ \mathrm{(terms)} & =- \frac{1}{12(1-x_{i})^{2}} \frac{Q^{2}}{M_{j}^{2}} \left( \frac{(1-x_{i})(3(1+x_{j}) -2x_{i}(2+6x_{j}-x_{j}^{2}) + x_{i}^{2}(1+x_{j}))}{(x_{i}-x_{j})(1-x_{j})^{2}} \right. \nonumber \\
    & \left. + \frac{(-6 + x_{i}(3 + 15x_{j} - 4x_{j}^{2}) + x_{i}^{2}(6-20x_{j}-4x_{j}^{2}) + x_{i}^{2}(6-20x_{j}+6x_{j}^{2}))\ln x_{j}}{(1-x_{j})^{3}} \right. \nonumber \\
    & \left. + \frac{x_{i}(3-4x_{j}+6x_{i}(1+x_{j})-12x_{i}^{2})\ln \left( \frac{x_{j}}{x_{i}} \right)}{(x_{i}-x_{j})^{2}} \right), \nonumber \\
    H^{h}(x_{i},x_{j};Q^{2}): \ \mathrm{(terms)} & = - \frac{Q^{2}}{M_{i}^{2}} \frac{x_{j}}{6(1-x_{i})^{2}} \left(- \frac{x_{j}(x_{i}(1+3x_{j})-(3+x_{j}))\ln x_{j}}{(1-x_{j})^{3}} \right. \nonumber \\
    & \left. + \frac{(1-x_{i})((1-3x_{j}-2x_{j}^{2})-x_{i}+5x_{i}x_{j})}{(x_{i}-x_{j})(1-x_{j})^{2}} + \frac{x_{j}(1-3x_{i})\ln \left( \frac{x_{j}}{x_{i}} \right)}{(x_{i}-x_{j})^{2}}\right).
    \label{eq7.1.44.5}
\end{align}
}

The $f_{B_{d}}$ and $f_{B_{u}}$ functions yield 
\begin{align}
        f_{B_{d}}(x_{i},y_{j}) & =  \left( 1 + \frac{1}{4} \frac{y_{j}}{x_{i}} \right)\Bar{d}^{lh}_{0}(x_{i},y_{j}) - 2\frac{y_{j}}{x_{i}}d^{lh}_{0}(x_{i},y_{j}),
    \label{eq7.1.47}
\end{align}
\begin{equation}
    f_{B_{u}}(x_{i},y_{j}) = - \left(4 + \frac{y_{j}}{4x_{i}} \right) \Bar{d}^{lh}_{0}(x_{i},y_{j}) + 2\frac{y_{j}}{x_{i}}d^{lh}_{0}(x_{i},y_{j}).
    \label{eq7.1.48}
\end{equation}
where $d_{0}^{lh}(x_{i},y_{j})$ and $\Bar{d}_{0}^{lh}(x_{i},y_{j})$ become
\begin{equation}
    d_{0}^{lh}(x_{i},y_{j}) = \frac{x_{i}^{2} \ \mathrm{ln}x_{i}}{(1-x_{i})^{2}(1-x_{i}y_{j})} + \frac{y_{j}x_{i} \  \mathrm{ln}y_{j}}{(1-y_{j})^{2}(1-x_{i}y_{j})} + \frac{x_{i}}{(1-x_{i})(1-y_{j})},
    \label{eq4.3.7}
\end{equation}
\begin{equation}
    \Bar{d}_{0}^{lh}(x_{i},y_{j}) = \frac{x_{i} \ \mathrm{ln}x_{i}}{(1-x_{i})^{2}(1-x_{i}y_{j})} + \frac{x_{i}(y_{j})^{2} \ \mathrm{ln}y_{j}}{(1-y_{j})^{2}(1-x_{i}y_{j})} + \frac{x_{i}}{(1-x_{i})(1-y_{j})}.
    \label{eq4.3.8}
\end{equation}

\section{Appendix: Hadronization tools }
\label{app:C}

Within the R$\chi$T framework, bilinear light quark operators coupled to the external sources are added to the massless QCD Lagrangian:
\begin{equation}
    \mathcal{L}_{\mathrm{QCD}} = \mathcal{L}_{\mathrm{QCD}}^{0} + \overline{q} \left[ \gamma_{\mu} (\upsilon^{\mu}+\gamma_{5}a^{\mu}) - (s-ip\gamma_{5}) \right]q,
    \label{e1}
\end{equation}
where the auxiliary fields defined as $\upsilon^{\mu} = \upsilon^{\mu}_{i} \lambda^{i}/2$, $a^{\mu} = a^{\mu}_{i} \lambda^{i}/2$, and $s = s_{i}\lambda^{i}$, with $\lambda^{i}$ the Gell-Mann matrices, are Hermitian matrices in  flavour space. Once the R$\chi$T action is fixed through $\mathcal{L}_{\mathrm{R}\chi\mathrm{T}}$, we can  hadronize the bilinear quark currents by taking the appropriate functional derivative with respect to the external fields:
\begin{align}
    S^{i} & = -\overline{q}\lambda^{i}q = \left.\frac{\partial \mathcal{L}_{\mathrm{R}\chi\mathrm{T}}}{\partial s_{i}} \right|_{j=0}, &\quad P^{i} & = \overline{q}i\gamma_{5}\lambda^{i}q = \left.\frac{\partial \mathcal{L}_{\mathrm{R}\chi\mathrm{T}}}{\partial p_{i}} \right|_{j=0}, \nonumber \\
    V^{i}_{\mu} & = \overline{q}\gamma_{\mu}\frac{\lambda^{i}}{2}q = \left.\frac{\partial \mathcal{L}_{\mathrm{R}\chi\mathrm{T}}}{\partial \upsilon_{i}^{\mu}} \right|_{j=0}, &\quad A^{i}_{\mu} & =  \overline{q}\gamma_{\mu}\gamma_{5}\frac{\lambda^{i}}{2}q = \left.\frac{\partial \mathcal{L}_{\mathrm{R}\chi\mathrm{T}}}{\partial a^{\mu}_{i}} \right|_{j=0},
    \label{e2}
\end{align}
where $j=0$ indicates that all external currents are set to zero.\\
The vector form factor from the $\gamma$ contribution to the decay into two pseudoscalar mesons is driven by the electromagnetic current
\begin{equation}
    V_{\mu}^{\mathrm{em}} = \sum_{d}^{u,d,s} Q_{q}\overline{q}\gamma_{\mu}q = V_{\mu}^{3}+ \frac{1}{\sqrt{3}}V_{\mu}^{8} = J_{\mu}^{\mathrm{em}},
    \label{e3}
\end{equation}
where $Q_{q}$ is the electric charge of the $q$ quark. We get also
\begin{align}
    \overline{u}\gamma_{\mu}P_{L}u & = J_{\mu}^{3} + \frac{1}{\sqrt{3}} V_{\mu}^{8} + \frac{2}{\sqrt{6}}J_{\mu}^{0}, \nonumber \\
    \overline{d}\gamma_{\mu}P_{L}d & = -J_{\mu}^{3} + \frac{1}{\sqrt{3}} V_{\mu}^{8} + \frac{2}{\sqrt{6}}J_{\mu}^{0}, \nonumber \\
    \overline{s}\gamma_{\mu}P_{L}s & = - \frac{2}{\sqrt{3}} V_{\mu}^{8} + \frac{2}{\sqrt{6}}J_{\mu}^{0},
    \label{e4}
\end{align}
where $J_{\mu}^{i} = (V_{\mu}^{i}-A_{\mu}^{i})/2$. The vector current contributes to an even number of pseudoscalar mesons or a vector resonance, while axial-vector current gives an odd number of pseudoscalar mesons.\\
In $Z$ contribution both vector and axial-vector currents do contribute:
\begin{align}
    J_{\mu}^{Z} & = V_{\mu}^{Z} + A_{\mu}^{Z}, \nonumber \\
    V_{\mu}^{Z} & = \frac{g}{2 c_{W}} \overline{q} \gamma_{\mu} \left[ 2 s_{W}^{2} Q_{q} - T_{3}^{(q)} \right] q, \nonumber \\
    A_{\mu}^{Z} & = \frac{g}{2c_{W}} \overline{q} \gamma_{\mu}\gamma_{5}T_{3}^{(q)}q,
    \label{e5}
\end{align}
with $Q_{q}$ (see eq. (\ref{eq7.1.4})) and $T_{3}^{(q)} = \mathrm{diag}(1,-1,-1)/2$ the electric charge and weak hypercharges, respectively.

\subsection{\texorpdfstring{$\tau \to \mu P$}{}}
The $\tau \to \mu P$ decay, in our model, is mediated only by axial-vector current ($Z$ gauge boson), as  $\mathcal{M}_{\gamma}$ does not contribute. The total amplitude for $\tau \to \mu P$ reads 
\begin{equation}
    \mathcal{M}_{\tau \to \mu P} = \mathcal{M}_{Z}^{P} + \mathcal{M}_{\mathrm{Box}}^{P},
    \label{e6}
\end{equation}
where $\mathcal{M}_{Z}^{P}$ is given by 

\begin{align}
        \mathcal{M}_{Z}^{P} & = -i \frac{g^{2}}{2c_{W}} \frac{F}{M_{Z}^{2}} C(P) \sum_{j} V_{\ell}^{j\mu*}V_{\ell}^{j\tau} \overline{\mu}(p')\left[\slashed{Q} (F_{L}^{Z}P_{L} + F_{R}^{Z}P_{R})\right]\tau(p), \nonumber \\
        \mathcal{M}_{\mathrm{Box}}^{P} & = -i g^{2} F \sum_{j} V_{\ell}^{j\mu*}V_{\ell}^{j\tau} B_{j}(P) \overline{\mu}(p^{\prime}) \left[\slashed{Q}P_{L} \right]\tau(p),
\label{e7}
\end{align}

where $Z_{L} = \frac{g}{c_{W}} (T_{3}^{q} - s_{W}^{2}Q_{q})$ and $Z_{R} = -\frac{g}{c_{W}}s_{W}^{2}Q_{q}$. The hadronization of the quark bilinear in $\mathcal{M}_{Z}$ is determined by vector and axial-vector currents from eq. (\ref{e5}), which are written in terms of one $P$ meson, turning out to be \cite{Arganda:2008jj}
\begin{align}
    V_{\mu}^{Z} & = 0, \nonumber \\
    A_{\mu}^{Z} & = - \frac{g}{2c_{W}} F \{ C(\pi^{0})\partial_{\mu}\pi^{0} + C(\eta)\partial_{\mu}\eta + C(\eta')\partial_{\mu}\eta' \},
    \label{e8}
\end{align}
where $F \simeq 0.0922$ GeV is the decay constant of the pion and the $C(P)$ functions are given by \cite{Arganda:2008jj,Husek:2020fru}
\begin{align}
    C(\pi^{0}) & = 1, \nonumber \\
    C(\eta) & = \frac{1}{\sqrt{6}} \left( \sin{\theta}_{\eta} + \sqrt{2} \cos{\theta}_{\eta} \right), \nonumber \\
    C(\eta') & = \frac{1}{\sqrt{6}} \left( \sqrt{2}\sin{\theta}_{\eta} - \cos{\theta}_{\eta} \right).
    \label{e9}
\end{align}
The box amplitude is composed by the following $B_{j}(P)$ factors \cite{Lami} 
\begin{align}
    B_{j}(\pi^{0}) & = \frac{1}{2} (B_{d}^{j} - B_{u}^{j}), \nonumber\\
    B_{j}(\eta) & = \frac{1}{2\sqrt{3}} \left[ (\sqrt{2} \sin{\theta_{\eta}}-\cos{\theta_{\eta}})B_{u}^{j} + (2\sqrt{2} \sin{\theta_{\eta}}+\cos{\theta_{\eta}})B_{d}^{j} \right], \nonumber \\
    B_{j}(\eta') & = \frac{1}{2\sqrt{3}} \left[ ( \sin{\theta_{\eta}}-2\sqrt{2}\cos{\theta_{\eta}})B_{d}^{j} - (\sin{\theta_{\eta}}+\sqrt{2}\cos{\theta_{\eta}})B_{u}^{j} \right],
    \label{e10}
\end{align}
where the $B_{q}^{j}$ functions are the form factors from box diagrams and the angle $\theta_{\eta} \simeq -18^{\circ}$.\\
The branching ratio reads
\begin{equation}
    \mathrm{Br}(\tau \to \mu P) = \frac{1}{4\pi} \frac{\lambda^{1/2}(m_{\tau}^{2},m_{\mu}^{2},m_{P}^{2})}{m_{\tau}^{2}\Gamma_{\tau}} \frac{1}{2}\sum_{i,f} |\mathcal{M}_{\tau \to \mu P}|^{2},
    \label{e11}
\end{equation}
where $\Gamma_{\tau} \approx 2.267\times10^{-12}$ GeV and $\lambda(x,y,z) = (x+y-z)^{2} - 4xy$, thus $\mathcal{M}_{\tau \to \mu P}$ is given by eq. (\ref{e6}). Thus,
\begin{equation}
    \sum_{i,f} |\mathcal{M}_{\tau \to \mu P}|^{2} = \frac{1}{2m_{\tau}} \sum_{k,l} \left[ (m_{\tau}^{2}+m_{\mu}^{2}-m_{P}^{2})(a_{P}^{k}a_{P}^{l*}+b_{P}^{k}b_{P}^{l*}) + 2m_{\mu}m_{\tau}(a_{P}^{k}a_{P}^{l*}-b_{P}^{k}b_{P}^{l*})\right],
    \label{e12}
\end{equation}
with $k,l = Z,B$. Defining $\Delta_{\tau \mu} = m_{\tau}-m_{\mu}$ and $\Sigma_{\tau \mu} = m_{\tau}+m_{\mu}$ we get
\begin{align}
    a_{P}^{Z} & = - \frac{g^{2}}{2 c_{W}} \frac{F}{2} \frac{C(P)}{M_{Z}^{2}} \Delta_{\tau\mu}(F_{L}^{Z}+F_{R}^{Z}), \nonumber \\
    b_{P}^{Z} & = \frac{g^{2}}{2 c_{W}} \frac{F}{2} \frac{C(P)}{M_{Z}^{2}} \Sigma_{\tau\mu}(F_{R}^{Z}-F_{L}^{Z}), \nonumber \\
    a_{P}^{B} & = - \frac{g^{2}F}{2} \Delta_{\tau\mu} B_{j}(P), \nonumber \\
    b_{P}^{B} & = - \frac{g^{2}F}{2} \Sigma_{\tau\mu} B_{j}(P).
    \label{e13}
\end{align}

\subsection{\texorpdfstring{$\tau \to \mu PP$}{}}
These channels are mediated by $\gamma-$, $Z-$penguins and box diagrams. Using the the electromagnetic current (eq. (\ref{e3})), the electromagnetic form factor reads
\begin{equation}
    \langle P_{1}(p_{1})P_{2}(p_{2}) | V_{\mu}^{\mathrm{em}} | 0 \rangle = (p_{1}-p_{2})_{\mu}F_{V}^{P_{1}P_{2}}(Q^{2}),
    \label{e14}
\end{equation}
where $Q=p_{1}+p_{2}$ and $F_{V}^{P_{1}P_{2}}(Q^{2})$ is steered by both $I = 1$ and $I=0$ vector resonances. Then, the complete amplitude is given by
\begin{equation}
    \mathcal{M}_{\tau \to \mu PP} = \mathcal{M}_{\gamma}^{P\overline{P}} + \mathcal{M}_{Z}^{P\overline{P}} + \mathcal{M}_{\mathrm{Box}}^{P\overline{P}}.
    \label{e15}
\end{equation}
The next step is to hadronize the quark bilinears appearing in each amplitude. They turn out to be
\begin{align}
    \mathcal{M}_{\gamma}^{P\overline{P}} = & \frac{e^{2}}{Q^{2}} F_{V}^{P_{1}P_{2}}(s) \times \nonumber \\
    & \sum_{j} V_{\ell}^{j\mu*}V_{\ell}^{j\tau} \overline{\mu}(p') [Q^{2}(\slashed{p}_{q} -  \slashed{p}_{\overline{q}})F_{L}^{\gamma}(Q^{2})P_{L} + 2im_{\tau} p_{q}^{\mu}\sigma_{\mu\nu}p_{\overline{q}}^{\nu} F_{M}^{\gamma}(Q^{2})P_{R}]\tau(p), \nonumber \\
    \mathcal{M}_{Z}^{P\overline{P}} = & g^{2} \frac{2s_{W}^{2}-1}{2c_{W}M_{Z}^{2}} F_{V}^{P_{1}P_{2}}(s) \sum_{j} V_{\ell}^{j\mu*}V_{\ell}^{j\tau} \overline{\mu}(p')(\slashed{p}_{q} - \slashed{p}_{\overline{q}})[\gamma^{\mu} (F_{L}^{Z}P_{L} + F_{R}^{Z}P_{R})]\tau(p), \nonumber \\
    \mathcal{M}_{\mathrm{Box}}^{P\overline{P}} = & \frac{g^{2}}{2} F_{V}^{P_{1}P_{2}}(s) \sum_{j} V_{\ell}^{j\mu*}V_{\ell}^{j\tau} (B_{u}^{j} - B_{d}^{j})\overline{\mu}(p') (\slashed{p}_{q} - \slashed{p}_{\overline{q}}) P_{L} \tau(p).
    \label{e16}
\end{align}
After computing each amplitude, we get the following branching ratio
\begin{equation}
    \mathrm{Br} (\tau \to \mu PP) = \frac{\kappa_{PP}}{64 \pi^{3} m_{\tau}^{2} \Gamma_{\tau}} \int_{s_{-}}^{s_{+}} ds \int_{t_{-}}^{t_{+}} dt \frac{1}{2} \sum_{i,f}|\mathcal{M}_{\tau \to \mu PP}|^{2},
    \label{e17}
\end{equation}
where $\kappa_{PP}$ is $1$ for $PP = \pi^{+}\pi^{-}, K^{+}K^{-}, K^{0}\Bar{K}^{0}$ and $1/2$ for $PP = \pi^{0}\pi^{0}$ (we neglect this one as it vanishes in the charge conjugation symmetry limit). In terms of the momenta of the particles participating in the process, $s = (p_{q}+p_{\overline{q}})^{2}$ and $t = (p - p_{\overline{q}})^{2}$, so that 
\begin{align}
    t_{-}^{+} & = \frac{1}{4s} \left[ \left(m_{\tau}^{2} - m_{\mu}^{2}\right)^{2} - \left(\lambda^{1/2}(s,m_{P_{1}}^{2},m_{P_{2}}^{2}) \mp \lambda^{1/2}(m_{\tau}^{2},s,m_{\mu}^{2}) \right)^{2} \right], \nonumber \\
    s_{-} & = 4m_{P}^{2}, \nonumber \\
    s_{+} & = (m_{\tau}-m_{\mu})^{2}.
    \label{e18}
\end{align}

\subsection{\texorpdfstring{$\tau \to \mu V$}{}}
For these cases the branching ratio of $\tau \to \mu V$ is related with the $\tau \to \mu PP$ one by trying to implement the experimental procedure as follows
\begin{equation}
    \mathrm{Br}(\tau \to \mu V) = \left. \sum_{P_{1},P_{2}} \mathrm{Br}(\tau\to\mu P_{1}P_{2})\right|_{V}. 
    \label{e19}
\end{equation}
In the above equation the $s$ limits are now restricted to
\begin{equation}
    s_{-} = M_{V}^{2} - \frac{1}{2}M_{V}\Gamma_{V}, \quad s_{+} = M_{V}^{2} + \frac{1}{2}M_{V}\Gamma_{V}.
    \label{e20}
\end{equation}
Therefore, when $V = \rho, \phi$,  their branching ratios are given by
\begin{align}
    \mathrm{Br}(\tau \to \mu \rho) & = \left. \mathrm{Br}(\tau \to \mu \pi^{+}\pi^{-})\right|_{\rho}, \nonumber \\
    \mathrm{Br}(\tau \to \mu \phi) & = \left. \mathrm{Br}(\tau \to \mu K^{+}K^{-})\right|_{\phi} + \left. \mathrm{Br}(\tau \to \mu K^{0}\Bar{K}^{0})\right|_{\phi}.
    \label{e21}
\end{align}

\section{Appendix: Hadronic form factors }
\label{app:D}
The vector form factors $F^{PP}_{V}(s)$, defined by eq. (\ref{e14}) , are based on two key points:
\begin{itemize}
    \item At $s \ll M_{R}^{2}$ (being $M_{R}$ a generic resonance mass), the vector form factor should match the $\mathcal{O}(p^{4})$ result of $\chi$PT \cite{Gasser:1983yg, Gasser:1984gg}.
    \item Form factors of QCD currents should vanish for $s \gg M_{R}^{2}$ \cite{Lepage:1980fj}.
    \label{f1}
\end{itemize}
We include energy-dependent widths for the wider resonances $\rho(770)$ and $\rho(1450)$ or constant for the narrow ones: $\omega(782)$ and $\phi(1020)$. For the $\rho(770)$ we take the definition put forward in \cite{GomezDumm:2000fz}
\begin{align}
    \Gamma_{\rho}(s) = \frac{M_{\rho}s}{96\pi F^{2}} \left[ \sigma_{\pi}^{3}(s)\theta(s-4m_{\pi}^{2}) + \frac{1}{2}\sigma_{K}^{3}(s)\theta(s-4m_{K}^{2}) \right],
    \label{f2}
\end{align}
where $\sigma_{P}(s) = \sqrt{1 - 4 \frac{m_{P}^{2}}{s}}$, while $\rho(1450)$ is parameterized as follows \cite{Arganda:2008jj} 
\begin{align}
    \Gamma_{\rho'}(s) = \Gamma_{\rho'}(M_{\rho'}^{2}) \frac{s}{M_{\rho'}^{2}} \left( \frac{\sigma_{\pi}^{3}(s) + \frac{1}{2}\sigma_{K}^{3}(s)\theta(s-4m_{K}^{2})}{\sigma_{\pi}^{3}(M_{\rho'}^{2}) + \frac{1}{2}\sigma_{K}^{3}(M_{\rho'}^{2})\theta(s-4m_{K}^{2})} \right)\theta(s-4m_{\pi}^{2}),
    \label{f3}
\end{align}
with $\Gamma_{\rho'}(M_{\rho'}^{2}) = 400 \pm 60$ MeV \cite{PDG}. We get the following expressions for the vector form factors
\begin{align}
    F_{V}^{\pi\pi}(s) = & F(s) \exp{\left[ 2 Re\left( \Tilde{H}_{\pi\pi} (s)\right) + Re\left( \Tilde{H}_{KK}(s)\right) \right]},
    \label{f4}
\end{align}
\begin{align}
    F(s) = & \frac{M_{\rho}^{2}}{M_{\rho}^{2}-s-iM_{\rho}\Gamma_{\rho}(s)} \left[ 1 + \left( \delta \frac{M_{\omega}^{2}}{M_{\rho}^{2}} - \gamma \frac{s}{M_{\rho}^{2}} \right) \frac{s}{M_{\omega}^{2}-s-iM_{\omega}\Gamma_{\omega}} \right] - \frac{\gamma s}{M_{\rho'}^{2}-s-iM_{\rho'}\Gamma_{\rho'}(s)},
    \label{f5}
\end{align}
\begin{align}
    F_{V}^{K^{+}K^{-}}(s) = & \frac{1}{2}\frac{M_{\rho}^{2}}{M_{\rho}^{2}-s-iM_{\rho}\Gamma_{\rho}(s)}\exp{\left[ 2 Re\left( \Tilde{H}_{\pi\pi} (s)\right) + Re\left( \Tilde{H}_{KK}(s)\right) \right]} \nonumber \\
    & + \frac{1}{2} \left[ \sin^{2}{\theta_{V}}\frac{M_{\omega}^{2}}{M_{\omega}^{2}-s-iM_{\omega}\Gamma_{\omega}} + \cos^{2}{\theta_{V}}\frac{M_{\phi}^{2}}{M_{\phi}^{2}-s-iM_{\phi}\Gamma_{\phi}}  \right] \exp{\left[ 3Re \left( \Tilde{H}_{KK}(s) \right) \right]},
    \label{f6}
\end{align}
\begin{align}
    F_{V}^{K^{0}\overline{K^{0}}}(s) = & -\frac{1}{2}\frac{M_{\rho}^{2}}{M_{\rho}^{2}-s-iM_{\rho}\Gamma_{\rho}(s)}\exp{\left[ 2 Re\left( \Tilde{H}_{\pi\pi} (s)\right) + Re\left( \Tilde{H}_{KK}(s)\right) \right]} \nonumber \\
    & + \frac{1}{2} \left[ \sin^{2}{\theta_{V}}\frac{M_{\omega}^{2}}{M_{\omega}^{2}-s-iM_{\omega}\Gamma_{\omega}} + \cos^{2}{\theta_{V}}\frac{M_{\phi}^{2}}{M_{\phi}^{2}-s-iM_{\phi}\Gamma_{\phi}}  \right] \exp{\left[ 3Re \left( \Tilde{H}_{KK}(s) \right) \right]},
    \label{f7}
\end{align}
where we have defined the following terms
\begin{align}
    \beta & = \frac{\Theta_{\rho\omega}}{3M_{\rho}^{2}}, \nonumber \\
    \gamma & = \frac{F_{V}G_{V}}{F^{2}}(1+\beta)-1, \nonumber\\
    \delta & = \frac{F_{V}G_{V}}{F^{2}}-1,\nonumber\\
    \Tilde{H}_{PP}(s) & = \frac{s}{F^{2}}M_{P}(s),\nonumber\\
    M_{P}(s) & = \frac{1}{12} \left( 1-4\frac{m_{P}^{2}}{s} \right)J_{P}(s) - \frac{k_{P}(M_{\rho})}{6} + \frac{1}{288\pi^{2}},\nonumber\\
    J_{P}(s) & = \frac{1}{16\pi^{2}} \left[ \sigma_{P}(s)\ln \frac{\sigma_{P}(s)-1}{\sigma_{P}(s)+1} + 2 \right],\nonumber\\
    k_{P}(\mu) & = \frac{1}{32\pi^{2}} \left( \ln \frac{m_{P}^{2}}{\mu^{2}}+1 \right).
    \label{f8}
\end{align}
The $\beta$ parameter includes the contribution of the isospin breaking $\rho - \omega$ mixing through $\Theta_{\rho\omega} = -3.3 \times 10^{-3}$ GeV$^{2}$ \cite{Pich}. The asymptotic constraint on the $N_{C} \to \infty$ vector form factor indicates $F_{V}G_{V} \simeq F^{2}$ \cite{GEcker}. We will use ideal mixing between the octet and singlet vector components, $\theta_{V} = 35^{\circ}$. We note that when isospin-breaking effects are turned off, the resummation of the real part of the chiral loop functions is not undertaken and the contribution from the $\rho'$ is neglected, the well-known results from the vector-meson dominance hypothesis are recovered. More elaborated form factors are obtained using the results presented here as seeds for the input phaseshift in the dispersive formulation, see e.g. refs. \cite{GomezDumm:2013sib, Gonzalez-Solis:2019iod}. These refinements modify only slightly the numerical results obtained with the form factors quoted in this appendix.



\end{document}